\documentclass[usenatbib,useAMS]{mnras}
\usepackage{graphicx}	
\usepackage{amsmath}	
\usepackage{amssymb}	
\usepackage{multicol}  
\usepackage{graphicx}
\usepackage{bm}
\usepackage{pdflscape}
\usepackage{multirow}
\usepackage{ulem}
\usepackage{longtable}
\usepackage[T1]{fontenc}
\usepackage{ae,aecompl}
\usepackage{newtxtext,newtxmath}

\title[Galaxy merger rates in $z = 2.24$ protoclusters]{\textbf{What boost galaxy mergers in two massive galaxy protoclusters at $z = 2.24$ }}

\author[S. Liu et al.]{Shuang~Liu,$^{1,2}$
Xian~Zhong~Zheng,$^{1,2}$\thanks{Contact e-mail: \href{mailto:xzzheng@pmo.ac.cn}{xzzheng@pmo.ac.cn}}, 
Dong~Dong~Shi,$^{1}$ 
Zheng~Cai,$^{3}$
Xiaohui~Fan,$^{4}$
Xin~Wang,$^{5,6,7}$ 
\and
Qirong~Yuan,$^{8}$
Haiguang~Xu,$^{9}$
Zhizheng~Pan,$^{1}$
Wenhao~Liu,$^{1}$
Jianbo~Qin,$^{1}$
Yuheng~Zhang$^{1,2}$
and Run~Wen$^{1,2}$
\\
$^{1}$Purple Mountain Observatory, Chinese Academy of Sciences, 10 Yuanhua Road, Nanjing, Jiangsu 210023,  China\\
$^{2}$School of Astronomy and Space Sciences, University of Science and Technology of China, Hefei, Anhui 230026, China\\
$^{3}$Department of Astronomy, Tsinghua University, Beijing 100084,  China\\
$^{4}$Steward Observatory, University of Arizona, 933 North Cherry Avenue, Tucson, AZ 85721, USA\\
$^{5}$School of Astronomy and Space Science, University of Chinese Academy of Sciences (UCAS), Beijing 100049, China\\
$^{6}$National Astronomical Observatories, Chinese Academy of Sciences, Beijing 100101, China\\
$^{7}$Institute for Frontiers in Astronomy and Astrophysics, Beijing Normal University, Beijing 102206, China\\
$^{8}$Department of Physics and Institute of Theoretical Physics, Nanjing Normal University, Nanjing 210023, China \\
$^{9}$School of Physics and Astronomy, Shanghai Jiao Tong University, 800 Dongchuan Road, Shanghai 200240, China
}
\date{\today}

\pubyear{2023}

\begin{document}
\label{firstpage}
\pagerange{\pageref{firstpage}--\pageref{lastpage}}
\maketitle

\begin{abstract} 
Characterizing the structural properties of galaxies in high-redshift protoclusters is key to our understanding of the environmental effects on galaxy evolution in the early stages of galaxy and structure formation.  In this study, we assess the structural properties of 85 and 87 H$\alpha$ emission-line candidates (HAEs) in the densest regions of two massive protoclusters, BOSS1244 and BOSS1542, respectively, using \textit{HST} $H$-band imaging data. Our results show a true pair fraction of $22\pm5 (33\pm 6$)\,per\,cent in BOSS1244 (BOSS1542), which yields a merger rate of $0.41\pm0.09$ ($0.52\pm0.04$)\,Gyr$^{-1}$ for massive HAEs with $\log (M_\ast/{\rm M}_\odot) \ge 10.3$. This rate is 1.8 (2.8) times higher than that of the general fields at the same epoch. 
Our sample of HAEs exhibits half-light radii and S\'ersic indices that cover a broader range than field star-forming galaxies. 
Additionally, about 15\,per\,cent of the HAEs are as compact as the most massive ($\log(M_{\ast}/{\rm M}_\odot) \gtrsim 11$) spheroid-dominated population. These results suggest that the high galaxy density and cold dynamical state (i.e., velocity dispersion of $<400$\,km\,s$^{-1}$) are key factors that drive galaxy mergers and promote structural evolution in the two protoclusters. Our findings also indicate that both the local environment (on group scales) and the global environment play essential roles in shaping galaxy morphologies in protoclusters. This is evident in the systematic differences observed in the structural properties of galaxies between BOSS1244 and BOSS1542.
\end{abstract}



\section{Introduction}\label{sec1:introduction}

Galaxies are known to reside in the cosmic web, which is composed of clusters, sheets, filaments, and voids. These structures are predominantly made up of cold dark matter and formed through the hierarchical merging of smaller structures over cosmic time \cite[]{Blumenthal+1984, Davis+1985, Cole+2000}. Understanding the formation and evolution of galaxies in different environments in connection with the assembly of large-scale structures is a central goal of modern astrophysics. While the evolution of galaxies in general fields has been extensively studied using current deep extragalactic surveys, recent efforts have focused on characterizing galaxy properties in overdense environments, such as galaxy clusters, at high redshifts.

Galaxy clusters are the largest gravitationally-bound systems in the Universe and are thought to accelerate the evolution of galaxies \citep{Thomas+2005}. At low redshifts, they are known to host most massive early-type galaxies with old stellar populations \cite[e.g.][]{Kodama+1998, Goto+2003, Mei+2009}. Conversely, at $z>\sim 2$, they are dominated by late-type star-forming galaxies (SFGs) in the early formation stage, known as protoclusters \cite[e.g.][]{Tanaka+2011, Andreon+2013, Overzier+2016, Alberts+2022}. The evolution in star formation and morphology for member galaxies of clusters is tightly coupled with the assembly processes of the clusters \cite[]{Dressler+1980, vanDokkum+2001, Poggianti+2006, Zirm+2012, Chan+2018}. However, a consistent picture of galaxy evolution in clusters, especially in massive protoclusters at $z>2$, is still lacking.

In particular, it remains unclear how the \textit{global} and \textit{local} environments influence the structural properties of galaxies in these overdense structures. It is not yet known whether the sizes of galaxies in protoclusters systematically differ from those in the general fields at the same epoch. The size of a galaxy is primarily regulated by its angular momentum gained from large-scale tidal torques and may be modified by mergers and secular processes \cite[e.g.][]{Jiang+2019}. In the overdense environment of protoclusters, a high galaxy density would significantly enhance the interaction effects and perhaps result in a higher merger rate. Therefore, dissecting the processes governing the size growth or compaction of galaxies in $z>2$ protoclusters is critical. 

At $z>2$, galaxies are believed to efficiently accrete gas through cold streams and form extended and clumpy discs \cite[]{Dekel+2009, Pillepich+2019, ForsterSchreiber+2020, Kretschmer+2020}. Observational studies on galaxy morphologies in $z>2$ protoclusters are limited, and it is unclear how the properties of galaxies in these environments differ from those in general fields. Answering these questions would address important issues related to galaxy formation and evolution in the context of large-scale structures. 

Galaxy mergers tend to occur more frequently in group environments with moderate velocity dispersions (a few hundred\,km\,s$^{-1}$), and are rare in local mature clusters with high velocity dispersions (500--1000\,km\,s$^{-1}$) \cite[]{McIntosh+2008, Tran+2008, Kocevski+2011}. It is expected that a higher fraction of mergers would occur in protoclusters. This has been confirmed by several studies on merger counts in $z\sim 1.5$--3 protoclusters \cite[e.g.][]{Lotz+2013, Hine+2016, Coogan+2018, Watson+2019}. However, some protoclusters at similar redshifts have not exhibited an enhancement in the merger fraction \cite[]{Delahaye+2017, Monson+2021}, likely due to differences in the protocluster's dynamical state. Pre-processing in groups before infalling into mature cores is also known to play a crucial role in shaping galaxy star formation and morphology \cite[]{Fujita+2004, Bianconi+2018, Kuchner+2022}.

Galaxy merging is an essential process in the formation of massive elliptical galaxies \cite[]{Naab+2014, Tadaki+2014, vanderWel+2014}, as well as triggering intense starbursts \cite[]{Mihos&Hernquist+1996}. The size growth of early- and late-type galaxies since $z=2$--3 \cite[e.g.][]{Daddi+2005, Trujillo+2006, Trujillo+2007, van_Dokkum+2008} is primarily driven by galaxy minor/major mergers in the general fields \cite[]{Bell+2005, van_Dokkum+2005, Bluck+2012, Newman+2012, Nipoti+2012}. In $z>2$ protoclusters, quiescent galaxies start to emerge \citep{Kubo+2021, McConachie+2022, Shi+2023} and appear larger in size than their field counterparts \citep[][but see also \citealt{Allen+2015}]{Lani+2013}. Meanwhile, most protocluster members at $z>2$ are still forming stars due to the large amounts of gas in the filamentary substructures \citep{Dekel+2009, Umehata+2019, Daddi+2021}. The effects of the environment on the size growth of SFGs in $z<2$ clusters were found to be inconsistent across different studies \citep[e.g.][]{Lotz+2013, Bassett+2013, Perez-Martinez+2021}. The study of stellar mass-size relation in comparison with field and cluster environments reveals that the sizes of cluster SFGs may be larger \citep{Lani+2013, Afanasiev+2023}, smaller \citep{Kuchner+2017}, or not significantly different \citep{Matharu+2019}. These inconsistencies observed in the cluster environment may be due to differences in the dynamical state of clusters, sample size, and selection bias related to stellar populations. Therefore, a detailed investigation of the morphologies of SFGs in protoclusters at $z> 2$ would provide insight into the environmental effects on the size growth of SFGs in the early Universe.

We have used the MApping the Most Massive Overdensity (MAMMOTH) technique \citep{Cai+2016, Cai+2017a, Cai+2017b} to identify two massive protoclusters, BOSS1244 and BOSS1542, at $z=2.24$. These were traced by H$\alpha$ emission-line galaxies over an area of $\sim400$\,arcmin$^2$ using narrow- and broad-band imaging \citep{Zheng+2021}, and spectroscopically confirmed as the most overdense protoclusters to date at $z>2$ \citep{Shi+2021}. With the high-spatial resolution of near-infrared imaging from \textit{HST}/WFC3, we are able to investigate the rest-frame optical morphologies of galaxies in the high-density regions of the two protoclusters, measure the size of regular galaxies, and select galaxy mergers to study the role of mergers in driving the structural evolution.

This paper is structured as follows. In Section~\ref{sec2:data}, we introduce the \textit{HST} observations and galaxy sample. In Section~\ref{sec3:results}, we present our main results regarding the merger rate and the stellar mass-size relationship. In Section~\ref{sec4:discussion}, we discuss our findings and their implications, and summarise our results in Section~\ref{sec5:summary}. Throughout this work, we adopt a cosmology with $\Omega_{m}=0.3$, $\Omega_{\Lambda}=0.7$, and $H_{0}=70$\,km$^{-1}$\,Mpc$^{-1}$. Magnitudes are given in the AB system \citep{Oke+1983} unless otherwise mentioned.

\section{Observations and Analysis}\label{sec2:data}

Using data obtained with the Canada-France-Hawaii Telescope (CFHT), \citet{Zheng+2021} identified 244 and 223 emitter-line objects in BOSS1244 and BOSS1542, respectively. These objects were identified by analyzing the deep near-infrared narrow-band H$2$S(1) and broad-band ($K{\rm s}$) imaging data with H$_2$S(1) $<$22.5\,mag, significant factor $\Sigma>3$, and rest-frame equivalent width $EW>45$\,\AA. It is expected that about 80\,per\,cent of emitters with an H$\alpha$ flux $>$2.5$ \times 10^{-17}$\,erg\,s$^{-1}$\,cm$^{-2}$ are HAEs at $z=2.24 \pm 0.021$ \citep{Zheng+2021}. The fraction is even higher in high-density regions.

The near-infrared spectroscopic observations with MMT and LBT confirmed 46 HAEs over $\sim 55$\,arcmin$^2$  in BOSS1244 and 36 HAEs over $\sim 61$\,arcmin$^2$ in BOSS1542. Density maps revealed that BOSS1244 has three components, and two of them project to form an elongated structure with a line-of-sight velocity dispersion of $405\pm 202$\,km\,s$^{-1}$ for the South-West at $z=2.230\pm 0.002$ and $377\pm 99$\,km\,s$^{-1}$ for the North-East  at $z=2.246\pm 0.001$. BOSS1542 is a giant filamentary structure with a line-of-sight velocity dispersion of $247\pm 32$\,km\,s$^{-1}$ \citep{Shi+2021}.

\subsection{\textit{HST} Observations}

\begin{figure*}
\begin{minipage}{0.495\linewidth}
  \centerline{\includegraphics[width=1\textwidth]{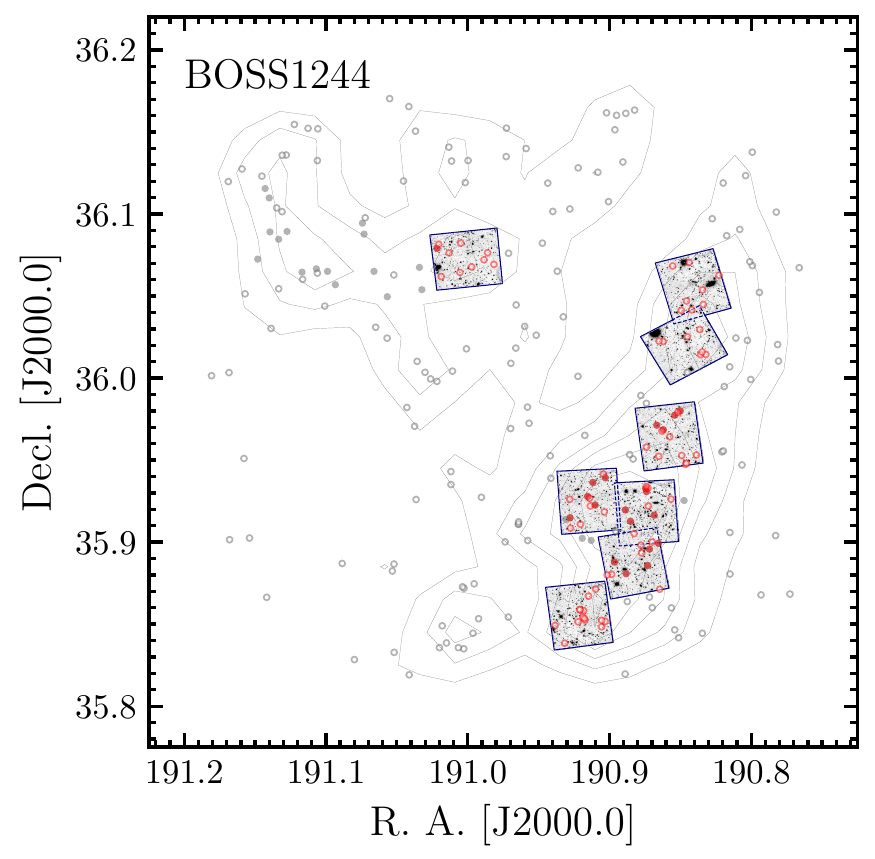}}
\end{minipage}
\begin{minipage}{0.495\linewidth}
  \centerline{\includegraphics[width=1\textwidth]{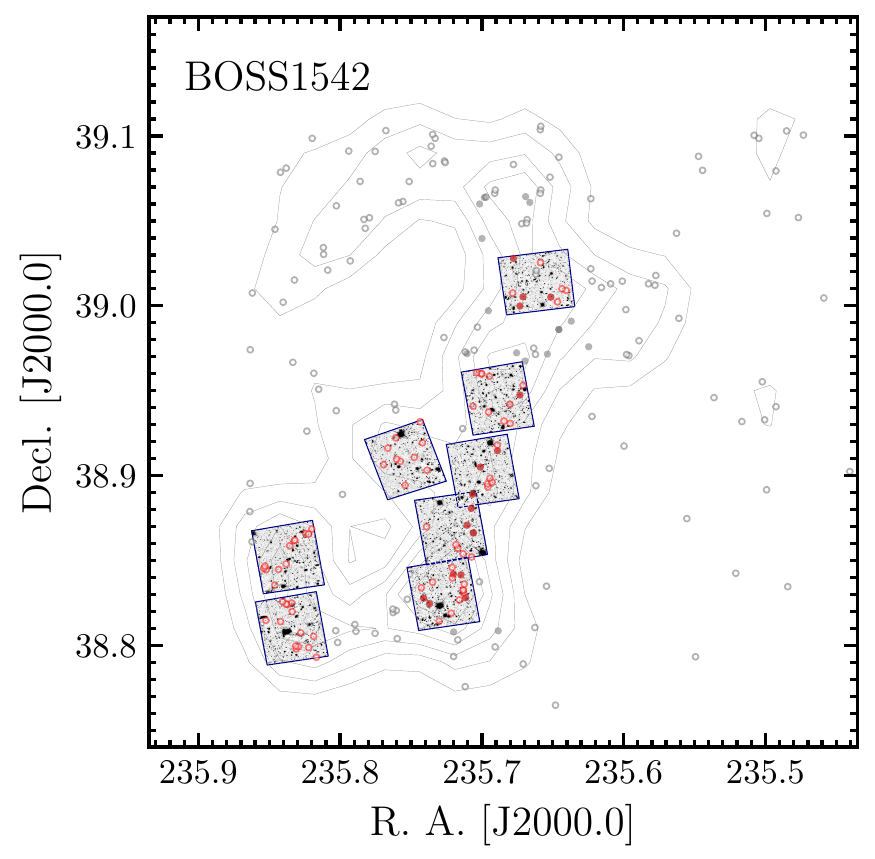}}
\end{minipage}
    \caption{Footprints of the \textit{HST}/WFC3 $H$-band imaging observations on the density maps of  HAEs at $z = 2.24$ in BOSS1244 (left) and BOSS1542 (right). Gray open circles represent the  HAEs from \protect\cite{Zheng+2021}, and gray filled circles mark the spectroscopically confirmed HAEs from \protect\cite{Shi+2021}. Red open and filled circles mark these  HAEs covered by the $H$-band observations.  The density maps and contour levels are adopted from \protect\cite{Zheng+2021}, showing 4, 8, 12, 16, 20 and 24 times higher than the HAE density in the general fields (0.071\,arcmin$^{-2}$) smoothed with a Gaussian kernel of $\sigma = 1\arcmin$. } \label{fig1_hst}
\end{figure*}

The near-infrared imaging data were obtained using the Wide Field Camera-3 (WFC3) on board the \textit{Hubble Space Telescope} (\textit{HST}) through near-infrared filter F160W ($H$) centred at $\lambda_{\rm p}=1.573$\,$\micron$ to a one-orbit depth (GO proposal 15266, PI: Z.~Cai). Eight pointings were targeted on the peak densities in the BOSS1244 and BOSS1542 fields, with each pointing covering $2.1\times 2.3$\,arcmin$^2$ and an effective exposure time of 2614\,s. The footprints of these pointings are shown in Fig.~\ref{fig1_hst}. The $H$-band imaging data were processed using the data reduction software tool {\tt{Astrodrizzle}}, and the final science images were produced with a scale of $0\farcs06$\,pixel$^{-1}$.

Out of 244 and 223 HAEs in BOSS1244 and BOSS1542, respectively, 85 and 87 HAEs are covered by the \textit{HST}/WFC3 $H$-band imaging observations. Of these, 20 and 16 have spectroscopic redshifts at $z\sim 2.2$ \citep{Shi+2021}. The average HAE density in the regions covered by the \textit{HST} observations in BOSS1244 and BOSS1542 is estimated to be $2.23\pm 0.07$\,arcmin$^{-2}$, which is about 30 times that of the general fields to the same detection depths \citep[0.071\,arcmin$^{-2}$;][]{An+2014,Zheng+2021}. In this study, we refer to HAEs without redshift confirmation as HAEs for simplicity, as the fraction of non-HAEs is low ($<\sim 10$\,per\,cent).

\subsection{Analysis of Galaxy Morphologies}\label{galft fitting}

We utilized the {\textsc{SExtractor}} software \citep{Bertin+1996} and {\textsc{Galfit}} \citep{Peng+2002, Peng+2010} embedded in {\textsc{Galapagos}} \citep{Barden+2012, Vika+2013} to perform various tasks, such as source detection, photometry, and galaxy two-dimensional S\'ersic model fitting on the $H$-band images of 172 HAEs in our two protoclusters. Following procedures described in \cite{Barden+2012}, we fitted a S\'ersic model to the $H$-band science image of each sample galaxy, and obtained the best-fitting values for the effective radius $r_{\rm e}$ (encloses 50\,per\,cent of the light along the major axis), the S\'ersic index $n$, and the total $H$-band flux.

For HAEs in BOSS1244 and BOSS1542, we estimated their stellar masses using $K_{\rm s}$- and $z$-band photometry. The observed $z-K_{\rm s}$ color corresponds to the rest-frame NUV\,$-R$ color for star-forming galaxies (SFGs) at $z=2.24$, which we used as a probe of the light-to-mass ratio ($L/M$). We derived the stellar mass from the $K_{\rm s}$ luminosity (the rest-frame $R$) and $z-K_{\rm s}$ for our HAEs at $z=2.24$ using the formulae (3) and (6) provided in \cite{Koyama+2013}. We obtained stellar masses for 76 HAEs in BOSS1244 and 75 HAEs in BOSS1542. However, some HAEs in BOSS1244 and BOSS1542 did not have $z$-band photometry, so we excluded them from our analysis involving stellar mass.

\section{Results}\label{sec3:results}

\subsection{Identification of Galaxy Mergers and Close Pairs}

A merger event between two galaxies can be broken down into four stages, namely (1) approaching each other; (2) passing the pericentre; (3) undergoing a damped oscillation while losing energy and angular momentum through dynamical friction; and (4) merging violently into the final coalescence. These stages are characterized by different timescales and morphologies that depend on the orbital parameters of the merging systems \citep{Lotz+2008,Lotz+2010}. Signs commonly used to recognize mergers include close pairs, double nuclei, or disturbed morphologies featuring tidal features. However, uncertainties still exist in identifying galaxy mergers \citep{Watson+2019}. For example, detecting the disturbed morphologies of a merging system heavily depends on image depth \citep{Ji+2014}. To address these uncertainties and the depth of imaging in two protoclusters, we visually classify galaxies with disturbed morphologies and identify close pairs to select merger candidates in different stages.

\begin{table*}
\centering
  \caption{Fractions of merging galaxies and close pairs among HAEs with $\log(M_\ast/{\rm M}_\odot) \geq 10.3$ in BOSS1244 and BOSS1542 in comparison with the general fields. $\Re_{\rm pair}$ and $\Re_{\rm merger}$ are calculated through pair fraction and merger fraction, respectively.}\label{tab1}
  \begin{tabular}{cccccccccc}
  \hline\hline
  {\rm Field} & {\rm Sample} & $N_{\rm tot}$ & $N_{\rm pair}$ & $N_{\rm merger}$  &$f_{\rm pair}$ (\%) &  $f_{\rm pair,corr}$ (\%) & $f_{\rm merger}$ (\%) &  $\Re_{\rm pair}$ ({\rm Gyr}$^{-1}$)  &  $\Re_{\rm merger}$ ({\rm Gyr}$^{-1}$) \\
  \hline
  \multirow{2}{*}{BOSS1244} & $z_{\rm spec}$ &  18   &  10    &  4  &  56$\pm$12   &  27$\pm$10   &  22$\pm$10   &  0.42$\pm$0.19   &  0.28$\pm$0.26     \\ 
                            &  all           &  61   &  31  &  16   &   51$\pm$6   &   22$\pm$5   &  26$\pm$6   &   0.41$\pm$0.09    & 0.33$\pm$0.24    \\
  \hline
  \multirow{2}{*}{BOSS1542} & $z_{\rm spec}$ &  13   &  12  &  4   &    92$\pm$8  & 63$\pm$13   &   31$\pm$13    &  0.96$\pm$0.27   & 0.39$\pm$0.36   \\ 
                            &   all          &  61   &  38  &  19  &  62$\pm$6  &  33$\pm$6     &  39$\pm$6  &   0.52$\pm$0.04     & 0.49$\pm$0.32    \\
  \hline
 CANDELS &  all   &  455   &  55  &  $\cdots$  &  $\cdots$   &  12$\pm$2   &   $\cdots$    &  0.19$\pm$0.05    &  $\cdots$     \\
 \hline
 \end{tabular}
\end{table*}

\begin{figure}
\begin{minipage}{0.48\textwidth}
\includegraphics[width=0.33\textwidth]{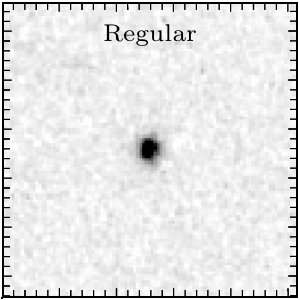}
\includegraphics[width=0.33\textwidth]{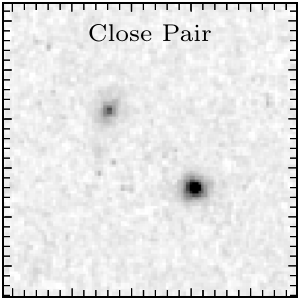}
\includegraphics[width=0.33\textwidth]{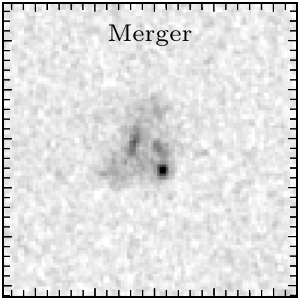}
\end{minipage}
\begin{minipage}{0.48\textwidth}
\includegraphics[width=0.33\textwidth]{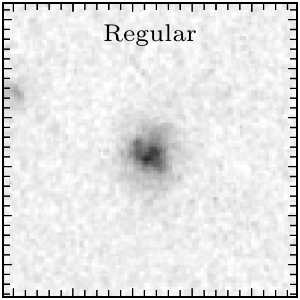}
\includegraphics[width=0.33\textwidth]{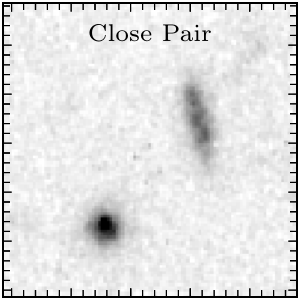}
\includegraphics[width=0.33\textwidth]{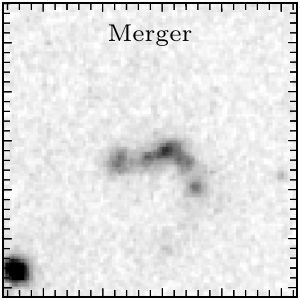}
\end{minipage}
    \caption{Example stamp images of HAEs of three morphological types  in our two protoclusters. These are \textit{HST}/WFC3 $H$-band  images in a size of $6\arcsec \times 6\arcsec$, corresponding $49.4\times 49.4$\,kpc$^2$ at $z=2.24$.} \label{fig2_examples}
\end{figure}

We classify the sample galaxies into three morphological types: (1) isolated regular galaxies with a smooth and symmetric appearance, (2) merging galaxies with disturbed morphologies, and (3) close pairs. We define a close pair as a major galaxy and a companion within a projected distance of $0\farcs61 < R_{\rm proj} < 3\farcs64$ (5--30\,kpc at $z=2.24$) with an $H$-band flux ratio of $\mu>0.25$. If multiple objects or galactic nuclei are distributed within a radius of $R_{\rm proj}\leq 0\farcs61$ ($\leq5$\,kpc), we refer such a system to as a merger. Example stamp images of the three morphological types of galaxies are presented in Fig.\ref{fig2_examples}. Fig.\ref{allimage} shows all \textit{HST} $H$-band images of the 172 HAEs in our sample. The $H$-band flux at $z\sim2$--3 corresponds to the rest-frame $B$-band and is mainly contributed by young stellar populations. Our $\mu>0.25$ cut is responsible for a lower stellar mass ratio cut because less massive galaxies tend to be younger and less attenuated by dust \citep{Man+2016}. This can introduce a bias in estimating the fraction of close pairs compared to the results based on the stellar mass ratio cut. We quantify this bias later using the control sample and correct our pair statistics accordingly.

More massive galaxies have a stronger level of clustering and tend to have a higher pair fraction \citep{OLeary+2021a}. To ensure a detection completeness above 80\,per\,cent, we limit our HAE sample to $\log(M_\ast/{\rm M}_\odot) \geq 10.3$, leaving 61 and 61 massive galaxies in BOSS1244 and BOSS1542, respectively, out of which 18 and 13 are spectroscopically confirmed. In total, we find 31 (38) close pairs and 16 (19) galaxies with disturbed and asymmetric morphologies for the massive HAEs in BOSS1244 (BOSS1542), giving an observed fraction of $51\pm 6$\,per\,cent ($62\pm 6$\,per\,cent) for close pairs and $26\pm 6$\,per\,cent ($39\pm 6$\,per\,cent) for merging systems (listed in Table~\ref{tab1}).

To make a comparison for filed and protocluster galaxies, we select a control sample from the 3D-HST/CANDELS catalogue to examine the fraction of close pairs and mergers in general fields. HAEs are representative of the main sequence of SFGs (refer to Fig. 8 in \citealt{An+2014}), and we choose SFGs that match the distribution of stellar mass and $K_{\rm s}$-band magnitude of protocluster HAEs, in order to reduce the differences between SFGs and HAEs. The control sample consists of 455 massive SFGs with $M_\ast \geq10^{10.3}$\,M$_\odot$ and $K{\rm s}<23.2$\,mag over $2.1<z<2.4$ (approximately equal to $2.24 \pm 3\sigma_z$, where $\sigma_z=0.06$ is the photometric redshift error in 68\,per\,cent confidence at $z>2.0$), matching the 80\,per\,cent completeness in stellar mass and $K_{\rm s}$-band brightness distribution of our HAE samples in BOSS1244 and BOSS1542.

To identify projected close pairs, we use the same criteria of $H$-band flux ratio $\mu>0.25$ and a projected distance of 5\,kpc\,$<R_{\rm proj}<$\,30\,kpc. The majority of galaxies in our control sample have photometric redshifts. Considering the uncertainties of photometric redshifts, we treat a galaxy pair with $\delta_z^2 < \sigma_{z,1}^2 + \sigma_{z,2}^2$ as a true pair, where $\sigma_{z,1}$ and $\sigma_{z,2}$ denote the photometric redshift uncertainties for the two galaxies in the pair \citep{Bundy+2009, Mantha+2018}. We determine the fraction of true close pairs for massive SFGs in general fields to be $12\pm 2$\,per\,cent.

\subsection{Corrections for False Pairs} \label{pair correction}

When identifying close pairs, it is common for us to lack spectroscopic redshifts for the companion galaxies. This can be due to foreground or background galaxies that are projected onto the same line of sight. The probability of encountering such contamination usually increases with galaxy density. In this study, we aim to quantify the probability of false close pairs resulting from such projections. 

To estimate the fraction of false pairs in the high-density environments of our two protoclusters, BOSS1244 and BOSS1542,  we employ a Monte Carlo simulation method. These protoclusters comprise 244 and 223 HAE candidates, respectively, covering a total area of approximately 400\,arcmin$^2$. Amongst these candidates, 80\,per\,cent (195 HAEs in BOSS1244 and 178 HAEs in BOSS1542) have been confirmed as true HAEs at $z\sim2.24$ \citep{Zheng+2021}. We utilize the COSMOS2020 catalog \citep{Weaver+2022}, which provides multi-wavelength deep observations over a sky area of approximately 1.29\,deg$^2$ in the COSMOS field, to calculate the likelihood of projected galaxy pairs, following the procedures below: 
(1) We randomly select a region of $20\times20$\,arcmin$^2$ from the COSMOS catalog and treat true COSMOS galaxies within this region as foreground and background objects.
(2) We randomly select 195 (178) SFGs with stellar masses $\log(M_\ast/{\rm M}_\odot) \geq 10.3$ and $K{\rm s}<23.2$\,mag over $2.1<z<2.4$ from the COSMOS field. These SFGs were chosen to match the $K_{\rm s}$ and stellar mass distributions of our HAE sample.
(3) We consider these selected SFGs as HAE analogues and randomly assign redshifts to them, taking into account the distribution of spectroscopic redshifts in BOSS1244 and BOSS1542 given in \cite{Shi+2021}.
(4) These analogues are then distributed within the chosen $20\times20$\,arcmin$^2$ region and are arranged to reproduce the density maps of BOSS1244 and BOSS1542 by being placed in between two contour lines in Fig.~\ref{fig1_hst}.
(5) We identify galaxy close pairs using our previous criteria and measure the fraction of projected pairs from foreground/background COSMOS field galaxies in the selected $20\times20$\,arcmin$^2$ region.
(6) To obtain the probability distribution of false pairs, we repeat procedures (1)--(5) one thousand times.

\begin{figure*}
\begin{minipage}{0.495\linewidth}
  \centerline{\includegraphics[width=1\textwidth]{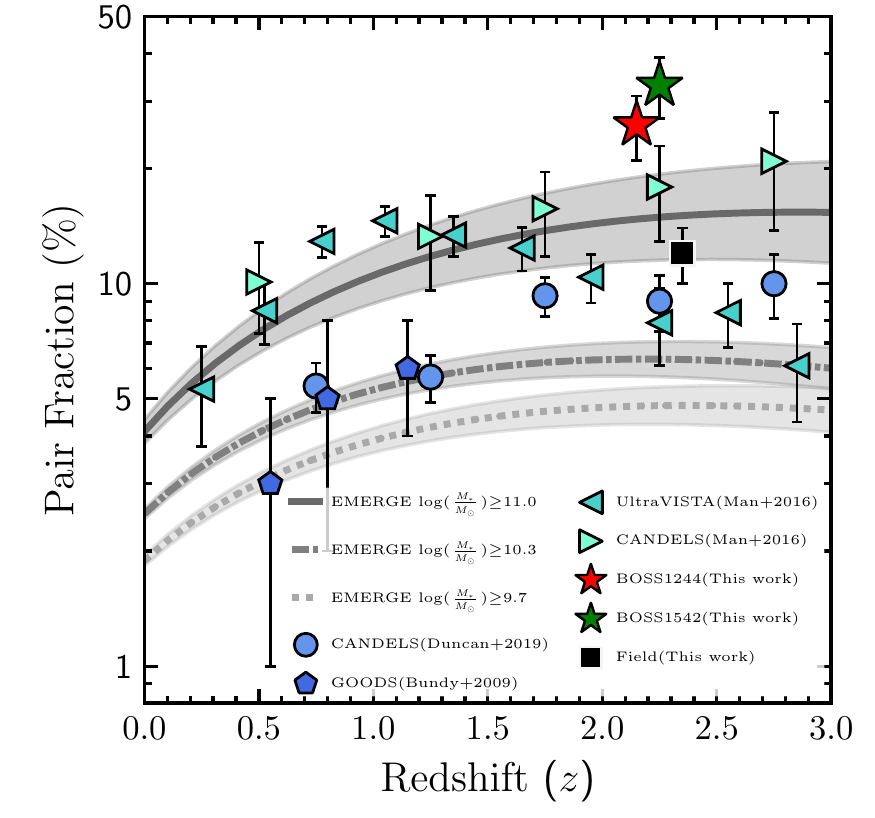}}
\end{minipage}
\begin{minipage}{0.495\linewidth}
  \centerline{\includegraphics[width=1\textwidth]{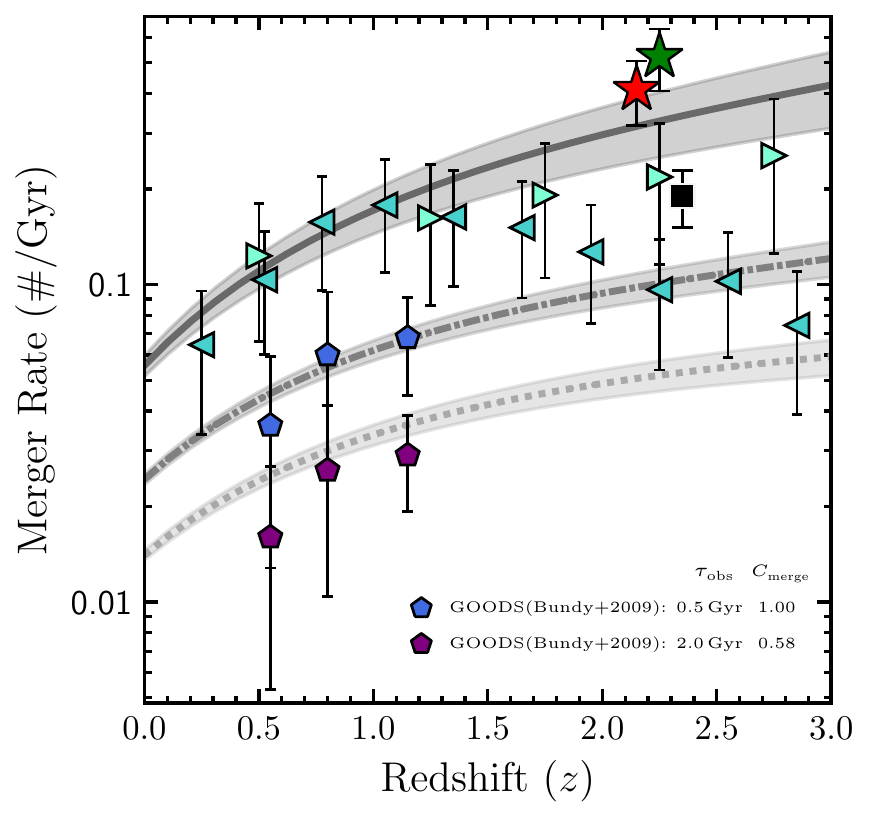}}
\end{minipage}
    \caption{Left: Pair fraction of galaxies as a function of redshift. The theoretical predictions from the {\textsc{Emerge}} simulation \protect\citep{OLeary+2021a} are denoted by the dotted, dashed-dotted, and solid lines for galaxies with $M_\ast \geq 10^{9.7}$\,M$_\odot$, $M_\ast \geq 10^{10.3}$\,M$_\odot$ and $M_\ast \geq 10^{11.0}$\,M$_\odot$, respectively. The gray shaded areas represent the Poisson error.  Data points refer to the observational estimates of close pair fraction from different  studies \protect\citep{Bundy+2009, Man+2016, Duncan+2019}.  The selection criteria used in these studies are presented in Section~\ref{pair correction}.  Our  close pairs are selected with projected distance of 5\,kpc $<R_{\rm proj}< 30$\,kpc and flux ratio of $\mu>0.25$ for galaxies with $M_\ast \geq 10^{10.3}$\,M$_\odot$ in BOSS1244 (red star), BOSS1542 (green star), and the general field (black square) at $z\sim 2.2$. The symbols are arbitrarily shifted in redshift for clarification.  Right: galaxy merger rate as a function of redshift. The shaded curves from the {\textsc{Emerge}} simulation are derived using a merger timescale $\tau_{\rm obs}$ linearly evolving with redshift. \protect\cite{Man+2016} estimated merger rates using $C_{\rm merge}=1$ and $\tau_{\rm obs}=0.81\pm 0.31$\,Gyr from simulations \protect\citep{Lotz+2010}. \protect\cite{Bundy+2009} adopted $C_{\rm merge} = 1$ and $\tau_{\rm obs} = 0.5$\,Gyr from \protect\cite{PA+2008} (blue diamonds),  and $C_{\rm merge}=0.58$ and $\tau_{\rm obs}=2$\,Gyr from \protect\cite{KW+2008} (purple diamonds) to deliver their estimates of galaxy merger rates. We take $\tau_{\rm obs} = 0.63\pm 0.05$\,Gyr and $C_{\rm merge}=1$ from the {\textsc{Emerge}} simulation to estimate the merger rates in BOSS1244, BOSS1542 and the general fields at $z\sim 2.2$.}\label{fig3_fpair}
\end{figure*}

Our analysis indicates that a considerable fraction of HAE analogues in the BOSS1244 (BOSS1542) protocluster, 32 (31)\,per\,cent are identified as close pairs. Among them, 3.1 (2.4)\,per\,cent are identified as true pairs of protocluster HAE analogues, implying that 28.9 (28.6)\,per\,cent are associated with false pairs. These false pairs are primarily contributed by foreground or background galaxies, while the variation in the number density of protocluster member galaxies has a minimal effect on the false-pair fraction, although the true-pair fraction tends to increase with an increasing number density of member galaxies. The observed close pair fraction for massive HAEs in BOSS1244 (BOSS1542) is $51\pm6$ ($62\pm6$)\,per\,cent. Based on our estimate, we determine that the fraction of true close pairs are approximately $22\pm5$ ($33\pm6$)\,per\,cent in BOSS1244 (BOSS1542). Our results suggest that the galaxy pair fraction in the BOSS1244 (BOSS1542) protocluster is 1.8 (2.8) times that of the general fields (i.e., ($12\pm2$)\,per\,cent), which are estimated using the same method. The left panel of Fig.~\ref{fig3_fpair} displays our estimates of pair fractions in the two protoclusters and the general fields, along with the close pair fractions obtained from the literature based on flux ratio \citep{Man+2016} and mass ratio \citep{Bundy+2009, Duncan+2019, OLeary+2021a}.

\cite{OLeary+2021a} identified companion galaxies within a projected annulus of $R_{\rm proj}=5$--30\,kpc for target massive galaxies with $\log(M_\ast/{\rm M}_\odot) \geq 9.7$, $\log(M_\ast/{\rm M}_\odot) \geq 10.3$, and $\log(M_\ast/{\rm M}_\odot) \geq 11$ from the {\textsc{Emerge}} simulation. Their results suggest that more massive galaxies contain a higher fraction of close pairs. \cite{Man+2016} identified close pairs using $R_{\rm proj}=14$--43\,kpc, $\log(M_\ast/{\rm M}_\odot) \geq 10.8$, and flux ratio $\mu>0.25$. It has been verified that the two distance cuts $R_{\rm proj} = 5$--30\,kpc and $R_{\rm proj} = 14$--43\,kpc yield similar counts of close pairs at $z>1.5$ \citep{Mantha+2018}. Although \cite{Man+2016} presented a higher close pair fraction at $z\sim 2.2$ based on CANDELS datasets than our general field estimate, we attribute the discrepancy mainly to the difference in the stellar mass cut. \cite{Duncan+2019} revisited close pairs in CANDELS using $\log(M_\ast/{\rm M}_\odot) \geq 10.3$, $R_{\rm proj} = 5$--30\,kpc, and $\mu_{\rm mass}>0.25$, giving a close pair fraction at $z\sim 2.2$ lower than our estimate. In \cite{Bundy+2009}, they used $R_{\rm proj}=$ 5--30\,kpc and $\mu_{\rm mass}>0.25$ to identify close pairs among massive galaxies with $\log(M_\ast/{\rm M}_\odot) \geq 10$. We emphasize that the final close pair fraction is sensitive to the selection criteria not only for the parent sample but also for galaxy pairs.

We also estimated the stellar masses of all companion galaxies in the identified close pairs to assess the effect on our results when changing the current flux ratio cut $\mu>0.25$ to a mass ratio cut $\mu_{\rm mass}>0.25$. Due to the limitations of our ground-based imaging data in terms of resolution and depth, we use \textit{HST}/WFC3 $J$-band (GO proposal 16276, PI: X.~Wang, \citealt{Wang+2022}) and $H$-band imaging data to estimate the stellar masses of all companion galaxies. The color between the two bands enables us to estimate the mass-to-light ratio of companion galaxies by matching galaxy spectral models from \cite{BC+2003} with an exponentially declining star formation history of $\tau = 3$\,Gyr, followed by the conversion of the $H$-band luminosity into stellar mass. Consequently, the adoption of a mass ratio cut of $\mu_{\rm mass}>0.25$ results in close pair fractions of $47\pm 6$\,per\,cent  and $51\pm 6$\,per\,cent for BOSS1244 and BOSS1542, respectively. Estimated close pair fractions were 10--20\,per\,cent lower than those based on the flux ratio cut. When we replace the flux ratio cut $\mu>0.25$ with a mass ratio cut $\mu_{\rm mass}>0.25$, we obtain a close pair fraction at $z \sim 2.2$ in the general fields consistent with that given in \cite{Duncan+2019}. However, the excess of close pair fractions in our two protoclusters compared to that in the general field remains unchanged within the uncertainties. We conclude that dense regions of BOSS1244 and BOSS1542 harbor close pairs of galaxies about 1.8 and 2.8 times that in the general field when the same criteria are applied.

\begin{figure*}
\begin{minipage}{0.4557\linewidth}
  \centerline{\includegraphics[width=1.0\textwidth,trim=0 0 30 0, clip]{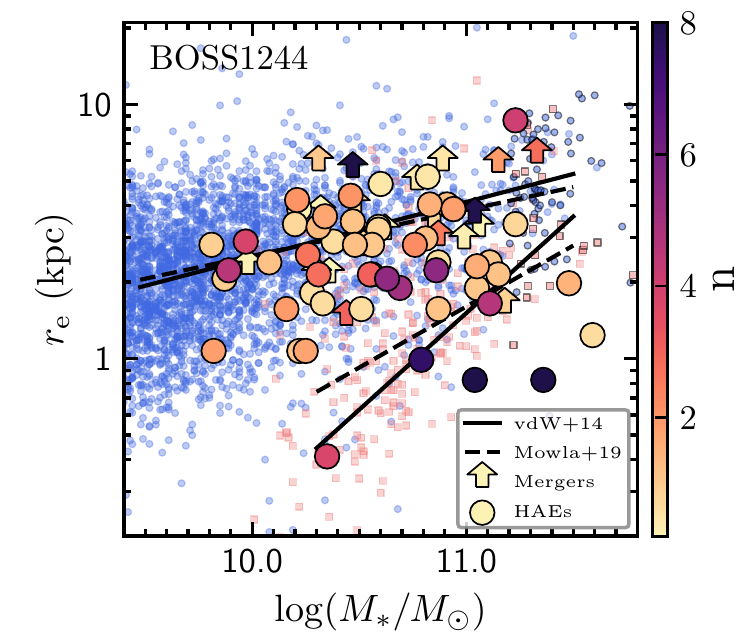}}
\end{minipage}
\begin{minipage}{0.5403\linewidth}
  \centerline{\includegraphics[width=0.88\textwidth,trim=23 0 0 0, clip]{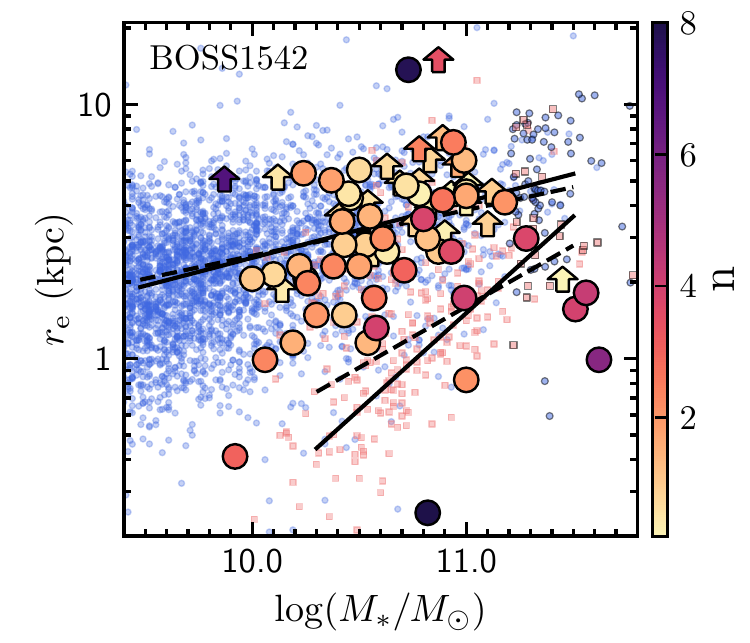}}
\end{minipage}
    \caption{Distributions of HAEs in the diagram of stellar mass versus half-light radius for BOSS1244 (left) and BOSS1542 (right).  The HAEs classified as close pairs and mergers (52 in BOSS1244 and 47 in BOSS1542) are denoted by filled circles, while filled upward arrows represent the lower limit of $r_{\rm e}$ for merging galaxies. The data points are colour coded by S$\acute{\rm e}$rsic index $n$. The late-type (small blue circles) and early-type (small red squares) galaxies at $2.0 \leq z \leq 2.5$ from 3D-HST/CANDELS are presented for comparison. The black solid lines represent the best-fitting relations for the two galaxy populations  given in \protect\cite{vanderWel+2014}.  Those symbols with black lines refer to massive galaxies detected in the COSMOS-DASH survey, and black dashed lines show the best-fitting relations to the combination of 3D-HST/CANDELS and COSMOS-DASH from \protect\cite{Mowla+2019}. } \label{fig4_msr}
\end{figure*}

We further estimate the projection probability of a companion galaxy within $R_{\rm proj}< 5$\,kpc using the same Monte Carlo method. We find that less than 1\,per\,cent of double-nuclei systems can be contaminated by foreground or background galaxies, indicating that merging systems with two galactic nuclei are marginally contaminated due to the projection effect. We noticed that about 20\,per\,cent of galaxies are both identified as {\textit{merger}} and {\textit{close pair}} categories in BOSS1542, indicating the presence of galaxy groups with multiple galaxies. In contrast, such systems are rare in BOSS1244. We hypothesize that this discrepancy in galaxy local density is determined by the global dynamic state of protoclusters --- BOSS1542 is dynamically cold with a velocity dispersion of $247\pm 32$\,km\,s$^{-1}$, while BOSS1244 is dynamically hot with two distinct components of $\sim 400$\,km\,s$^{-1}$ velocity dispersion each.

\subsection{Galaxy Merger Rate}

In our morphological classification, close pairs are considered to be galaxy mergers in the early approaching stages, with separation distances ranging from 5\,kpc to 30\,kpc, while galaxies with disturbed morphologies are considered to be ongoing major mergers in the final stage. Galaxy merger rates can be inferred from either pair fractions or merger fractions.

The merger rate per galaxy is defined by the equation:
\begin{equation}\label{model}
\Re [{\rm Gyr^{-1}}] = \frac{C_{\rm merge} \times f_{\rm pair}}{\tau_{\rm obs}},
\end{equation}
where $C_{\rm merge}$ refers to the correction factor for the possibility that galaxy pairs will merge together, and $\tau_{\rm obs}$ is the timescale of a merging event. The merging timescale has been evaluated through merger simulations \citep{Lotz+2010}, cosmological simulations \citep{KW+2008, Jiang+2014}, and dynamical friction  \citep{Bell+2006,Boylan-Kolchin+2008, Jiang+2008}. \cite{OLeary+2021a} compared the merging timescales from different studies with those from the \textsc{Emerge} simulation and found that a timescale linearly changing with redshift is best able to reproduce the intrinsic merger rate. We estimate the merger rates for protocluster galaxies using the merging timescale relation $\tau_{\rm obs}=w(1+z)+b$, where $w= -0.177\pm 0.007$ and $b=1.205\pm 0.023$ for massive galaxies with $\log(M_\ast/{\rm M}_\odot) \geq 10.3$ and major companion galaxies within 5\,kpc to 30\,kpc \citep{OLeary+2021a}. For $z=2.24$, the merging timescale is $\tau_{\rm obs} = 0.63\pm 0.05$\,Gyr. Adopting $C_{\rm merge}=1$ for our close pairs, we estimate the galaxy merger rate to be $0.41\pm 0.09$\,Gyr$^{-1}$ and $0.52\pm 0.04$\,Gyr$^{-1}$ for massive galaxies in BOSS1244 and BOSS1542, respectively.  We display our merger rates converted from the pair fractions for BOSS1244, BOSS1542, and general fields in the right panel of Fig.~\ref{fig3_fpair}. Our field merger rate at $z\sim 2.2$ is consistent with that given in \cite{Man+2016}, in which $C_{\rm merge}=1$ and $\tau_{\rm obs}=0.81\pm 0.31$\,Gyr are adopted \citep{Lotz+2010}.

Moreover, the galaxy merger rate can be estimated from the fraction of morphologically-identified merging galaxies. We emphasize that the recognition of morphological disruption in a galaxy image strongly depends on the image depth. From our one-orbit \textit{HST} $H$-band images of $z\sim 2.24$ HAEs, only the disruption signatures of high surface brightness are seen. This cosmic dimming effect certainly leads to a decrease in the number of merging galaxies. \cite{Lotz+2008} determined a merging timescale of 0.4--1.2\,Gyr for the morphologically-selected merging galaxies using the C.A.S. method. The timescale uncertainty depends on the orbital and initial conditions of a merging system. We adopt a moderate value of $\tau_{\rm obs}= 0.8\pm 0.4$\,Gyr to estimate the galaxy merger rate, giving $0.33\pm 0.24$\,Gyr$^{-1}$ and $0.49\pm 0.32$\,Gyr$^{-1}$ in BOSS1244 and BOSS1542, respectively, which are in good agreement with those derived from the close pair fractions within the uncertainties. Again, we note that the merger rate estimates based on the fractions of merging galaxies suffer from large uncertainties in identification as well as the merging timescale. We tabulate our estimates of the galaxy merger rates in Table~\ref{tab1}.

Fig.\ref{fig3_fpair} and Table\ref{tab1} clearly demonstrate that the merger rates of HAEs in both BOSS1244 and BOSS1542 are significantly higher than those of field SFGs at the same cosmic epoch. This indicates that during the rapid assembly of these protoclusters, violent interactions and merging processes were also occurring amongst galaxies. In order to fully comprehend galaxy evolution in overdense environments during cosmic noon, it is imperative to understand what is driving the enhancement of galaxy mergers in these protoclusters. Furthermore, we observe that BOSS1542 exhibits a one-quarter higher merger rate than BOSS1244. This reliable difference is not induced by shot noise.

In contrast, the subsample of spectroscopically-confirmed HAEs in BOSS1542 displays an even higher fraction of close pairs compared to the overall HAE sample, although such a trend is not evident in the spec-$z$ subsample of BOSS1244. We believe this may, in part, be due to selection bias, as the spectroscopic observations were heavily focused on the strongly clustered HAEs. Nevertheless, we infer from the disparity in merger rates between the two protoclusters that BOSS1542 appears to favor boosting galaxy mergers to a greater extent than BOSS1244.

\subsection{The Stellar Mass-Size Relation}

We analyze the stellar mass-size relation of our HAE sample in BOSS1244 and BOSS1542. Galaxy size corresponds to the half-light radius of the galaxy, measured from the \textit{HST} $H$-band image using {\textsc{Galfit}} S\'ersic profile fitting. As shown in Fig.~\ref{fig4_msr}, this diagram depicts the distribution of 76 HAEs in BOSS1244 on the left and 75 HAEs in BOSS1542 on the right, in the mass-size ($M_\ast$-$r_{\rm e}$) plane. The data points are colour-coded by their respective S\'ersic indices. Additionally, we select a mass-complete sample of SFGs with $2.0 \leq z \leq 2.5$ from 3D-HST/CANDELS and include them in Fig.~\ref{fig4_msr} for comparison. It is well-established that SFGs (late-type) and quiescent galaxies (early-type) follow two distinct scaling relations, as illustrated by the solid lines from \cite{vanderWel+2014} and the dashed lines from \cite{Mowla+2019}, which include the most massive galaxies ($M_\ast \geq 10^{11.3}$,M$_\odot$).

\begin{figure}
\begin{minipage}{0.48\textwidth}
\includegraphics[width=0.48\textwidth]{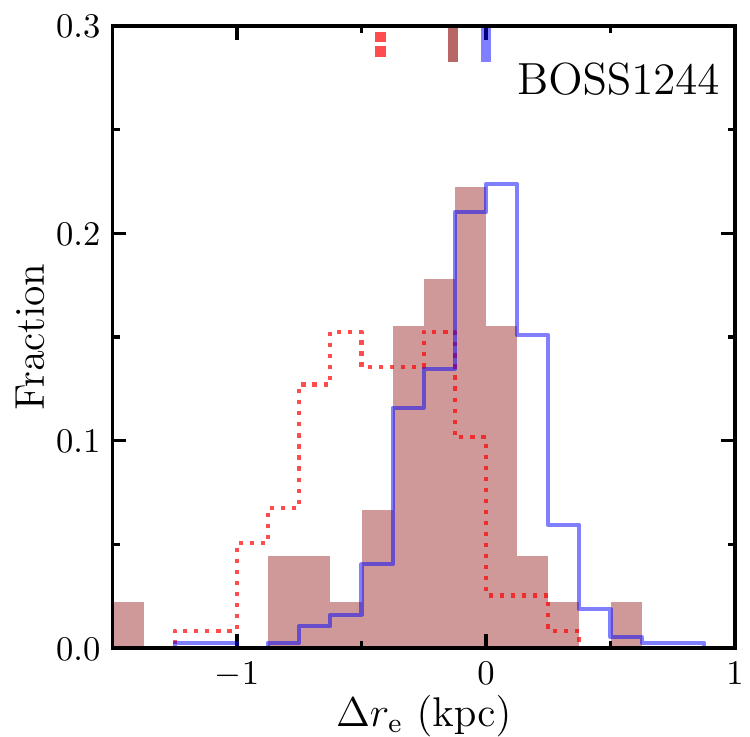}
\includegraphics[width=0.48\textwidth]{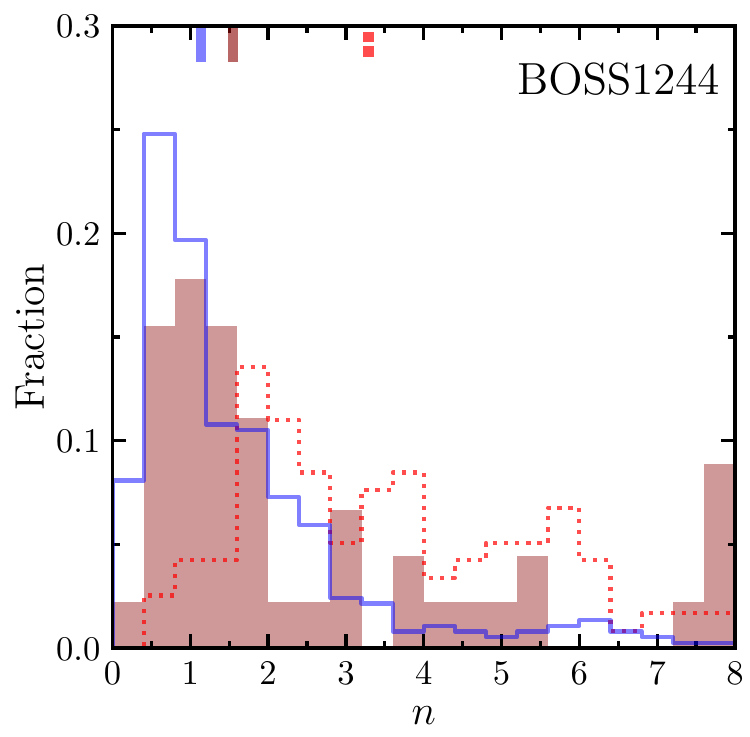}
\end{minipage}
\begin{minipage}{0.48\textwidth}
\includegraphics[width=0.48\textwidth]{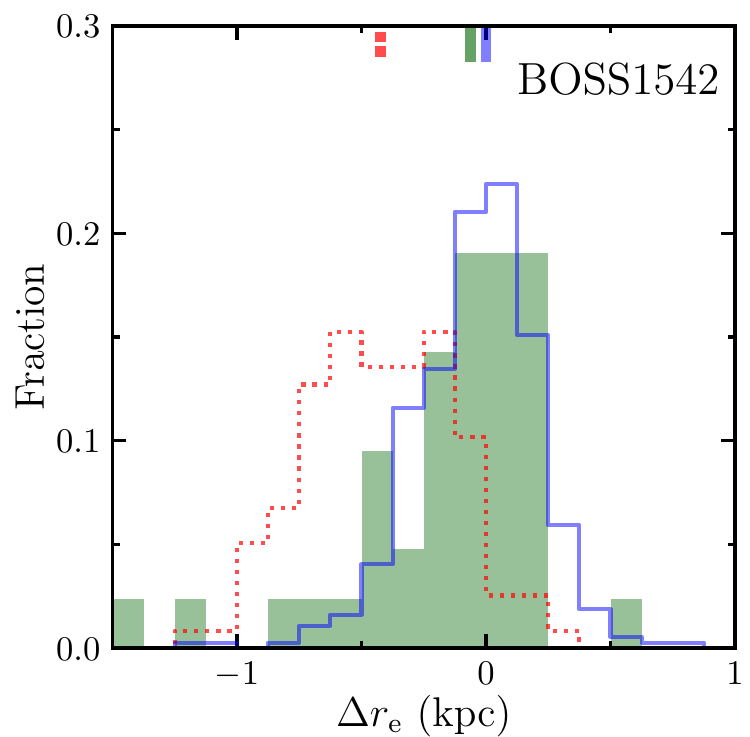}
\includegraphics[width=0.48\textwidth]{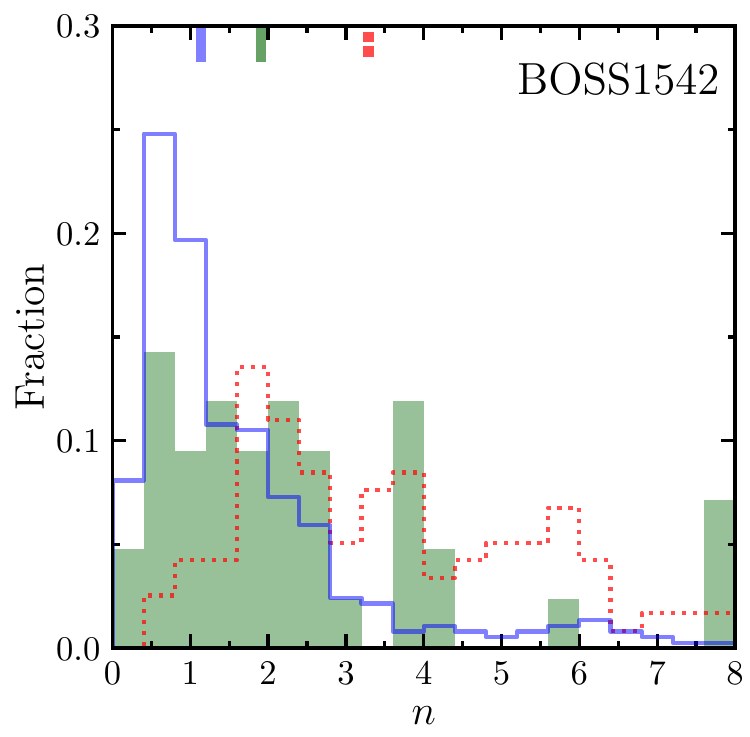}
\end{minipage}
    \caption{The distributions of size difference $\Delta r_{\rm e}$ and S$\acute{\rm e}$rsic index $n$ for HAEs with $M_{\ast} \geq 10^{10.3}$\,M$_\odot$ in BOSS1244 and BOSS1542, massive ($M_{\ast} \geq 10^{10.3}$\,M$_\odot$) SFGs and quiescent galaxies at $2.1<z<2.4$ in general field, are denoted by darked, green, blue and red histograms. The median values of each population are represented by short vertical lines in the top of each panel. The size difference \protect$\Delta r_{\rm e,i} = r_{\rm e,i}(M_{\ast,i}) - r_{\rm e_{best}, i}(M_\ast,i)$, where $r_{\rm e_{best}, i}(M_\ast,i)$ is the best-fitting $r_{\rm e}$ of SFGs given in \protect\cite{vanderWel+2014} at the given stellar mass of each target.} \label{fig5}
\end{figure}

We select our sample of $z=2.24$ HAEs such that most of them have stellar masses $\log(M_\ast/{\rm M}_\odot) \geq 10.3$. The majority of these HAEs follow the trend of field massive SFGs with increasing size at increasing stellar mass, as shown in Fig.~\ref{fig4_msr}. However, the protocluster HAEs exhibit distinct structural differences from the field massive SFGs. We demonstrate this statistically through the normalized distribution of size difference $\Delta r_{\rm e}$ and S\'ersic index ($n$) for HAEs in BOSS1244, BOSS1542, and SFGs/quiescent galaxies in the general field, as depicted in Fig.~\ref{fig5}.

We find that, unlike the trend followed by HAEs in the general field, these protocluster HAEs span both disc-dominated and spheroid-dominated regimes on the $M_\ast$-$r_{\rm e}$ diagram. Additionally, the scatter in size and S\'ersic index is larger in BOSS1244 and BOSS1542 compared to the general field massive SFGs. Moreover, using the two-dimensional Kolmogorov-Smirnov (K-S) test to compare structural parameters between HAEs in BOSS1244 (BOSS1542) and field SFGs, we find that the two sets of samples differ significantly with the $p$-value of 0.024 (0.055), as determined by the \textsc{ks2d2s} program \citep{Peacock+1983}. The median sizes of the HAEs in BOSS1244 (BOSS1542) are 2.80\,kpc (2.97\,kpc), which are smaller than that of the massive SFGs in the general field (3.27\,kpc). Additionally, the median S\'ersic indices of the HAEs in BOSS1244 (BOSS1542) is 1.55 (1.91), which are statistically larger than that of the field SFGs (median $n$ value of 1.13).

We identify nine (six) extremely massive ($\log(M_\ast/{\rm M}_\odot)\gtrsim11$) compact galaxies in BOSS1244 (BOSS1542), out of the 52 (47) HAEs classified as \textit{close pair $+$ regular}. These nine and six galaxies are undergoing active star formation (namely `blue nuggets'), and are expected to become `red nuggets' after their star-formation activity ceases \citep[e.g.][]{Zolotov+2015,Tacchella+2016}. Interestingly, we find that late-type galaxies dominate the compact galaxies in BOSS1244 whereas the opposite is true for compact galaxies in BOSS1542. 
The latter may have formed through major mergers. The exact processes that regulate the structural properties of these compact SFGs require further investigation beyond the scope of this work. However, the observed structural characteristics of the HAEs in the high-density regions of our two protoclusters are significantly influenced by the overdense environment and may be indicative of the protoclusters' dynamical states.

\section{Discussion}\label{sec4:discussion}

\subsection{what boost galaxy mergers in protoclusters at $z=$ 2--3?}

\subsubsection{Dynamical state of protoclusters}

The frequency of galaxy mergers is enhanced in environments of high galaxy density and low velocity dispersion, with a peak in galaxy groups in the local Universe \citep{Kocevski+2011}. However, mergers are not commonly seen in mature clusters due to their high velocity dispersion of 500--1000\,km\,s$^{-1}$. The galaxy merger rate increases with redshift out to at least $z \sim 2.5$, which is the peak epoch of the cosmic star formation density \citep[e.g.][]{OLeary+2021a}.

Galaxy protoclusters lie at the intersection of dense, gas-rich filaments in the cosmic web and gather galaxies and materials mostly from the surrounding filamentary structures into the central massive halo \citep{Ocvirk+2008,Overzier+2016}. The dynamical state of protoclusters is much controlled by surrounding tidal torque fields that impart the angular momentum of galaxies \citep{Ryden+1988}. In the early phase of assembly, protoclusters remain dynamically cold with velocity dispersion at a level of a few hundred\,km\,s$^{-1}$ before becoming dynamically hot with velocities greater than 500\,km\,s$^{-1}$. Therefore, it is expected that protoclusters with moderate velocity dispersion and high galaxy number density contain a high fraction of galaxy-galaxy encounters compared to the general field at the same epoch.

Our results support that dynamically-cold protoclusters contain a higher fraction of galaxy mergers. In the densest regions of the $z\sim 2.24$ protocluster BOSS1244 (BOSS1542), we obtained a merger rate of $0.41\pm0.09$ ($0.52\pm0.04$)\,Gyr$^{-1}$, respectively. This is 1.8 (2.8) times higher than that of the general field (i.e., $0.19\pm0.05$\,Gyr$^{-1}$), with the galaxy number density in these regions about 30 times higher than that of the general fields. BOSS1244 contains two substructures, each with a velocity dispersion of $\sim400$\,km\,s$^{-1}$, while BOSS1542 has a lower velocity dispersion of $\sim250$\,km\,s$^{-1}$ \citep{Shi+2021}. Interestingly, BOSS1542 is dynamically colder than BOSS1244 but exhibits a relatively higher galaxy merger rate. This observation confirms that the dynamical state of a galaxy protocluster is the key to boosting galaxy mergers.

Galaxy mergers have been shown to be enhanced in protoclusters based on previous studies. \cite{Lotz+2013} reported that the galaxy merger rate in the protocluster IRC-0218A at $z=1.62$ was 3--10 times higher compared to the general fields at the same epoch. This protocluster exhibits a high overdensity factor of 20 and a low velocity dispersion of $360\pm90$\,km\,s$^{-1}$ \citep{Pierre+2012}, being dynamically cold and therefore favorable for galaxy mergers. Conversely, \cite{Monson+2021} found no enhancement in the protocluster SSA22 at $z=3.09$ due to limited HST observations. SSA22 is comprised of two filamentary structures with velocity dispersions of 350\,km\,s$^{-1}$ and 540\,km\,s$^{-1}$, respectively \citep{Topping+2016}. The relatively low overdensity factors  ($9.5\pm2.0$ and $4.8\pm1.8$) of the two regions covered by the HST observations away from the density peak region of SSA22 \citep{Topping+2018}, suggests that it lacks the necessary conditions (overdensity and kinematics) to facilitate galaxy mergers. As expected, four clusters at $1.59<z<1.71$ with high velocity dispersions of $>500$\,km\,s$^{-1}$ does not show evidence of a merger enhancement in their central regions \citep{Delahaye+2017}. These findings suggest that not only global overdensity, but also local environments within the protocluster and/or overdensity, affect the rates of galaxy mergers.

\subsubsection{The local environment and preprocessing in the infall regions}

Galaxy groups represent the most efficient local environment for triggering galaxy mergers, as the potential well of a group halo binds member galaxies, dramatically increasing the probability of galaxy encounters. Conversely, galaxies commonly moving along filamentary structures are characterized by a lower merger frequency. Preprocessing, or physical processes in the group environment before entering massive cluster haloes, have been shown to play a role in this discrepancy \citep{Hough+2023}.
Hydrodynamical simulations indicate that galaxy groups may retain a relatively low velocity dispersion and facilitate galaxy mergers in the outskirts of a massive cluster \citep{Deger+2018,Benavides+2020}. The cluster tidal force may also dissolve about half of member galaxies in an infalling group by the first pericentre \citep{Haggar+2022}. In high-redshift protoclusters, preprocessing may therefore be a key factor in explaining the enhancement of galaxy merger rates \citep{Muldrew+2018,Chan+2021}. The fraction of infalling galaxies in galaxy groups and the dynamical state of protoclusters are expected to be the key factors in enhancing galaxy mergers in protoclusters.

It has been observed that a significant portion of member galaxies in local galaxy clusters (about 20-40\,per\,cent) were likely accreted as part of groups \citep{McGee+2009, Boselli+2022}. In the examined clusters up to $z\sim1.5$, nearly half of massive galaxies are found to be bounded by groups \citep{Boselli+2022}. The two massive protoclusters we study were undergoing rapid assembly processes. Our analysis shows that in the most dense regions of BOSS1542, the galaxies both identified as {\textit{merger} and {\textit{close pair} categories are more often associated with group-scale overdensities, whereas such systems are less frequently seen in BOSS1244. This supports our hypothesis that preprocessing plays a stronger role in raising the galaxy merger rate in BOSS1542 compared to BOSS1244. In contrast, BOSS1244 remains dynamically cold but more concentrated, and the effects of preprocessing are expected to become weaker as the two substructures merge and become a dynamically-hot cluster.

Furthermore, preprocessing has been shown to advance galaxy star formation and morphological transformation \citep{Oh+2018}. These effects have been reported in the infall regions of galaxy clusters at $0.2<z<1.5$ \citep{Bianconi+2018, Chan+2021} and reproduced in cosmological simulations \citep{Bahe+2013, Kuchner+2022}. Preprocessing in the infall regions of BOSS1542 could therefore strengthen star formation and reshape galaxy morphology through merging or interacting with other galaxies, or multiple fly-bys. Additionally, external perturbations outside the group halo could cause a loss of gas angular momentum, fueling star formation and even nuclear activities \citep{Boselli+2022}. This is supported by the excess of HAEs with a high H$\alpha$ luminosity observed in BOSS1542 \citep{Zheng+2021}.

\subsection{Effects of the overdense environment on galaxy structural evolution}\label{sec4.2:effect of env on size}

\subsubsection{Gas accretion and size growth}

The growth of a galactic disc is driven by the accretion of cooling gas within the dark matter halo and characterized by the approximately same specific angular momentum of the host halo. The halo acquires angular momentum from large-scale tidal torque, which is strongly dependent on halo mass and formation epoch. Therefore, the size of the disc is controlled by the transformation of gas that settles in the disc into stars from the halo \citep{Trujillo+2006,DeFelippis+2017}. As the gas collapses within the halo, the baryonic matter loses angular momentum via viscous redistribution \citep{Burkert+2009,Swinbank+2017} or gains angular momentum from strong galactic winds \citep{Shen+2003} in the regime between the virial radius and disc scale.

At $z\sim2$--3, overdense environments of protoclusters contain significant amounts of gas in their filamentary substructures \citep{Umehata+2019, Daddi+2021}. In these overdense regions, gas cold streams efficiently fuel star formation at a high rate, enabling the formation of relatively compact galactic discs \citep{Zolotov+2015, Oteo+2018}. The size growth of these discs could be compacted as long as the supplied gas is of low angular momentum. However, the tidal torque fields within protoclusters are complex, and the intensity of angular momentum and the transformation of angular momentum between baryonic components (stars, hot and cold gas) and dark matter are hardly outlined. It is, therefore, difficult to assess whether the $z\sim2$--3 protoclusters can make disc galaxies larger or smaller compared to the general fields. Our results show that the sizes of galaxies in BOSS1244 are noticeably smaller than those in the general fields in a statistically significant manner, while galaxies in BOSS1542 do not exhibit such a difference despite the relatively large scatters. We note that BOSS1244 is dynamically hotter than BOSS1542, and we lack solid evidence to link the discrepancy or consistency in galaxy size to the dynamical state of protoclusters or some physical processes in the overdense environments.

\subsubsection{Galaxy compaction}\label{sec4.2.2:galaxy compaction}
  
Galaxy (proto)clusters significantly increase the frequency of galaxy fly-bys, resulting in gravitational disturbances of galaxy discs and suppressed gas accretion. These disruptions may trigger disc instability and lead to a shrinkage of the discs. \cite{Matharu+2019} found that the size distribution of late-type cluster galaxies generally decreases with increasing velocity dispersion of the clusters, suggesting that fly-bys and interactions act as compaction processes in shrinking the size of galaxy discs. We speculate that such processes also occur in $z\sim2$--3 protoclusters, and the relatively smaller sizes of disc galaxies in BOSS1244 might be partially governed by gravitational compaction. Furthermore, we notice that the stellar mass-size relation of protocluster galaxies exhibits more scatter than that of the general fields. We argue that the increased scatter might be caused, in part, by the more frequent interruptions through fly-bys in the protocluster environments.

Multiple physical processes can cause galaxy compaction, including major mergers \citep{Kretschmer+2020,Tacconi+2020}, counter-rotating stream accretion, and disc instability \citep{DB+2014,Zolotov+2015}. Of these, disc instability and major mergers are considered as the major compaction mechanisms in protoclusters. At $z=2$--3, gas accretion from the cosmic web is so efficient that star-forming galaxies maintain a high fraction of gas, fueling star formation and even forming giant clumps. In this scenario, gaseous discs are fragile and easily collapse when triggered by instability \citep{Zolotov+2015}. Compact star-forming galaxies, also known as "blue nuggets," are thought to be regulated by rapid dissipative gas inflow and have short depletion times \citep{Tacchella+2016}. \cite{Husko+2022} pointed out that violent disc instability serves as a dominant channel for spheroid growth and starburst, while major mergers lead to violent dynamical processes that trigger starburst and size compaction.

We have observed the presence of extremely massive ($\log(M_{\ast}/{\rm M}_\odot) \gtrsim 11$) compact SFGs in the two protoclusters BOSS1244 and BOSS1542. 
Interestingly, the extremely massive compact SFGs in BOSS1542 are mostly spheroid-dominated with $n>2.5$, likely forming via mergers. These compact galaxies represent 29.5\,per\,cent of massive star-forming galaxies with $\log(M_{\ast}/{\rm M}_\odot) \geq 10.3$, comparable to the pair fraction of $33\pm6$\,per\,cent in BOSS1542. This consistence does not necessarily mean that all compact spheroid-dominated SFGs are merger remnants, as the collapse of gaseous discs may also form compact galaxies. We note that the compact massive SFGs in BOSS1244 are mostly disk-dominated with $n<2.5$, while the pair fraction is $26\pm5$\,per\,cent in BOSS1244. Given that BOSS1542 is dynamically colder than BOSS1244 and the latter is forming a dominant cluster core \citep{Shi+2023}, we argue that disk instability might be the dominant mechanism for the formation of massive compact SFGs in overdense environments, although major mergers may also form disc-dominant remnants \citep{Sotillo-Ramos+2022}. It is worthwhile to mention that the high merger rate measured in BOSS1244 and BOSS1542 has an observable impact on galaxy structural properties and is, in part, responsible for the increased scatter in the stellar mass-size relation in the two protoclusters.

\section{Summary}\label{sec5:summary}

We conducted a morphology analysis for H$\alpha$ emission-line candidates in two $z\sim2.24$ protoclusters, BOSS1244 and BOSS1542, using \textit{HST} $H$-band imaging data from 16 pointings covering the density peak regions. Our sample includes 85 out of 244 (87 out of 223) HAE candidates in BOSS1244 (BOSS1542), which were visually classified into three morphological types: regular, close pair, and merging galaxies. We derived their structural parameters using the two-dimensional S\'ersic profile fitting technique.

We found that the true close pair fraction in the protoclusters was higher compared to the general fields at $z=2$--3, with a true close pair fraction of $22\pm5 (33\pm 6$)\,per\,cent for BOSS1244 (BOSS1542). Monte Carlo simulations estimated a false pair fraction of 28.9 (28.6)\,per\,cent in BOSS1244 (BOSS1542) from projected foreground/background galaxies under the same selection criteria, which has been corrected in close pair fraction. We estimated the galaxy merger rate from the pair fraction, adopting a merging timescale of $0.63\pm0.05$\,Gyr, and obtained $0.41\pm0.09$ ($0.52\pm0.04$)\,Gyr$^{-1}$ for massive SFGs in the densest regions of BOSS1244 (BOSS1542), which were 1.8 (2.8) times that of the general fields with merger rate of $0.19\pm0.05$\,Gyr$^{-1}$ at the same epoch. Our estimation of merger rates from the fractions of morphologically-selected merging galaxies yielded consistent results, demonstrating an enhancement of galaxy merger rate in the two protoclusters. We argued that the cold dynamical state is the key factor in boosting the galaxy merger rate in the two massive overdense protoclusters, characterized by velocity dispersion of $< 400$\,km\,s$^{-1}$.

We found that our protocluster HAEs exhibit distinctive structural features compared to SFGs in general fields. Among the HAEs in our sample with $\log(M_\ast/{\rm M}_\odot) \geq 10.3$, the scatter in size (measured by half-light radius) is larger than that of field SFGs with similar stellar masses. In both BOSS1244 and BOSS1542, massive HAEs have a median size of 2.80\,kpc (2.97\,kpc), and are statistically smaller than the median size of 3.27\,kpc for massive field SFGs with $2<z<2.5$. The HAEs also exhibit a wider range of S\'ersic indices, with a higher median value (1.55 in BOSS1244 and 1.91 in BOSS1542) compared to field SFGs (1.13) which are mostly disc-dominated with $n<2.5$. 

Moreover, approximately 13-17\,per\,cent of these HAEs are as compact as the most massive ($\log(M{\ast}/{\rm M}_\odot) \gtrsim 11$) spheroid-dominated galaxies in the fields. Such large scatter in size and excess of compact HAEs demonstrate a significant impact of the overdense environment on galaxy structural evolution. We emphasize that the complex physical processes occurring in the dense regions of protoclusters, including efficient gas accretion, frequent mergers/interactions/fly-bys, and preprocessing in galaxy groups within infall regions, result in protocluster galaxies evolving more violent manner.

We pointed out that the structural properties of member galaxies are closely correlated with the dynamical state (or assembly stage) and density of protoclusters. For example, we found that BOSS1542 is dynamically colder than BOSS1244, containing more galaxy mergers and more compact HAEs with $n>2.5$. We argued that both local environment (over group scales) and global environment of protoclusters play crucial roles in shaping galaxy morphologies. Understanding the structural evolution of protocluster galaxies requires consideration of the preprocessing in infall regions and the assembly history.


\section*{Acknowledgements}

We thank the anonymous referee for her/his valuable comments and suggestions that improved this manuscript. 
This work is supported by the National Science Foundation of China (12233005, 12073078, 12173088, and 12273013), the science research grants from the China Manned Space Project with NO. CMS-CSST-2021-A02, CMS-CSST-2021-A04 and CMS-CSST-2021-A07, and the Chinese Academy of Sciences (CAS) through a China-Chile Joint Research Fund (CCJRF \#1809) administered by the CAS South America Centre for Astronomy (CASSACA).  XW is supported by CAS Project for Young Scientists in Basic Research, Grant No. YSBR-062. D.D.S. acknowledges support from China Postdoctoral Science Foundation (2021M703488), Jiangsu Funding Program for Excellent Postdoctoral Talent (2022ZB473) and Special Research Assistant Project  CAS.

\section*{Data Availability}

The data underlying this article are available in the article and in its online supplementary material.


\bibliographystyle{mnras}
\bibliography{ms_refs.bib}

\appendix
\section{Morphological parameters of HAEs in BOSS1244 and BOSS1542}

Table~\ref{alltable} presents the catalog of HAEs  in BOSS1244 and BOSS1542. The morphological parameters are derived with {\textsc{SExtractor}} and {\textsc{Galfit}} embedded within {\textsc{Galapagos}} from $H$-band mosaic images. In {\textsc{Galfit}}, the S\'ersic index is set between 0.2 and 8, the half-light radius $r_{\rm e}$ is set within 400 pixels, and magnitude is set between $-$3 and 3\,dex from the {\textsc{SExtractor}} magnitude.

\begin{figure*}
    \includegraphics{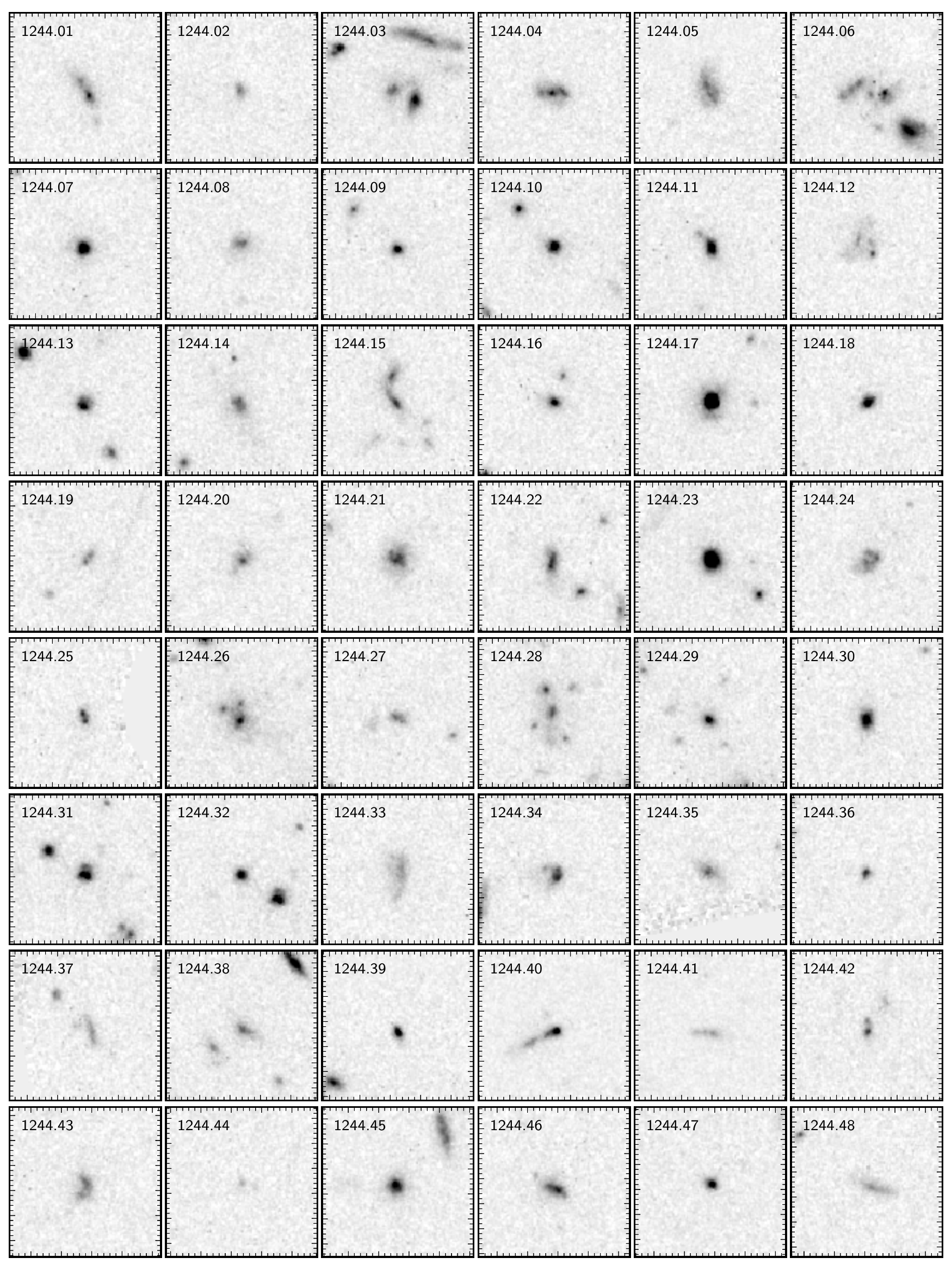}
    \caption{\textit{HST}/WFC3 $H$-band stamp images for 173 HAEs in BOSS1244/BOSS1542. Each stamp in a size of $6\arcsec \times 6\arcsec$ ($49.4\times 49.4$\,kpc$^2$ at $z = 2.24$).}
    \label{allimage}
\end{figure*}

\begin{figure*}
    \includegraphics{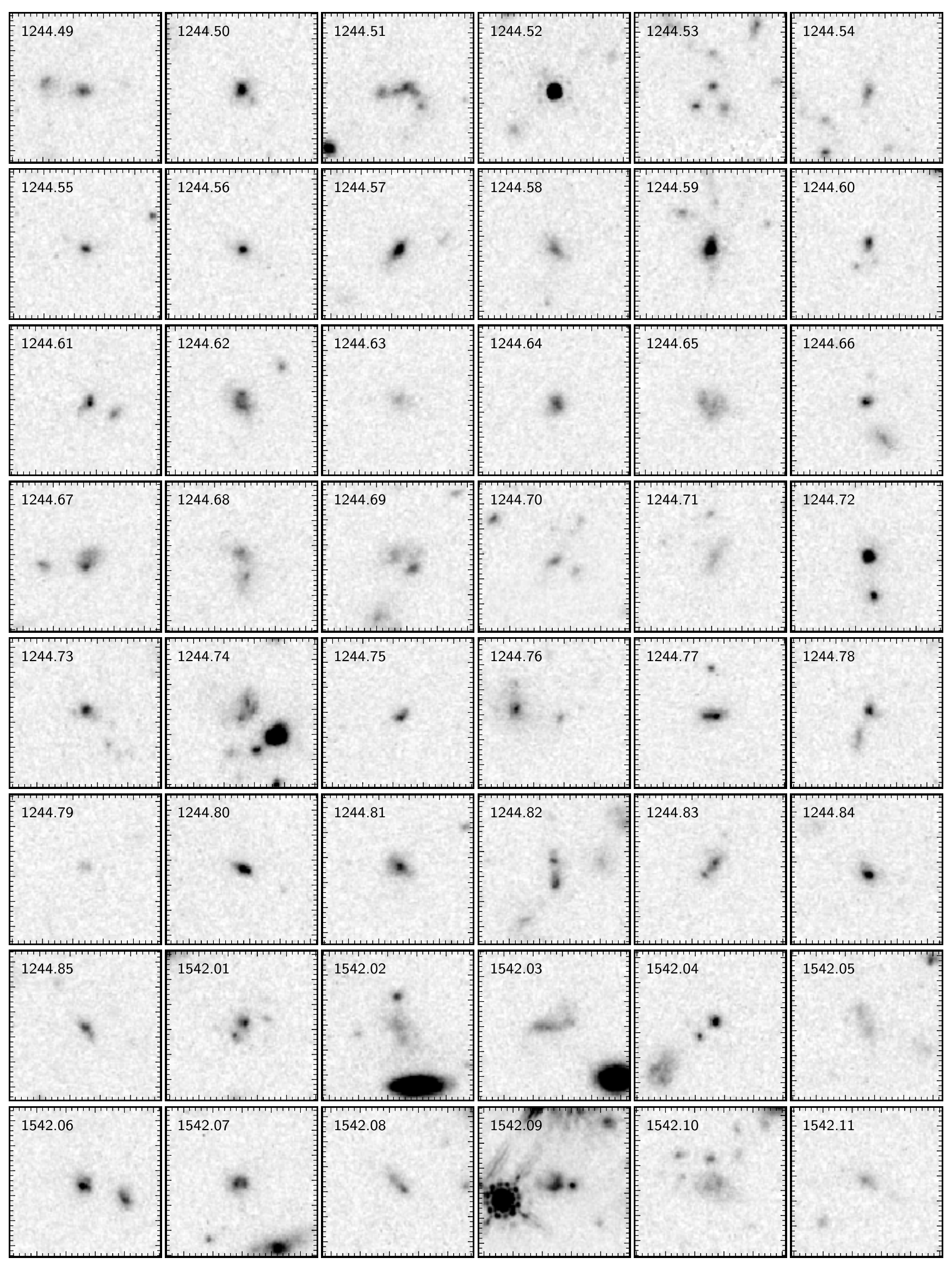}
    \contcaption{}
\end{figure*}

\begin{figure*}
    \includegraphics{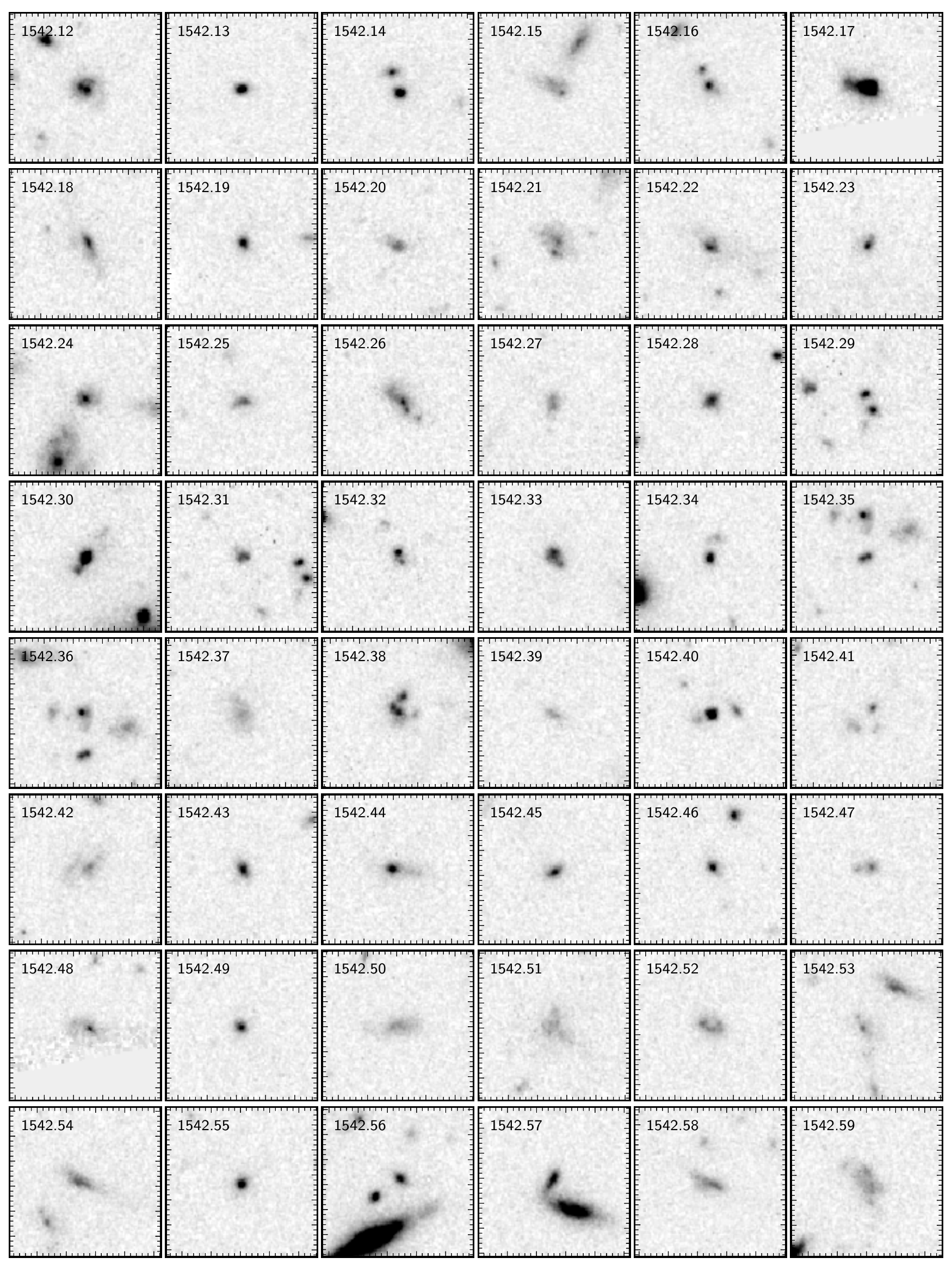}
    \contcaption{}
\end{figure*}

\begin{figure*}
    \includegraphics{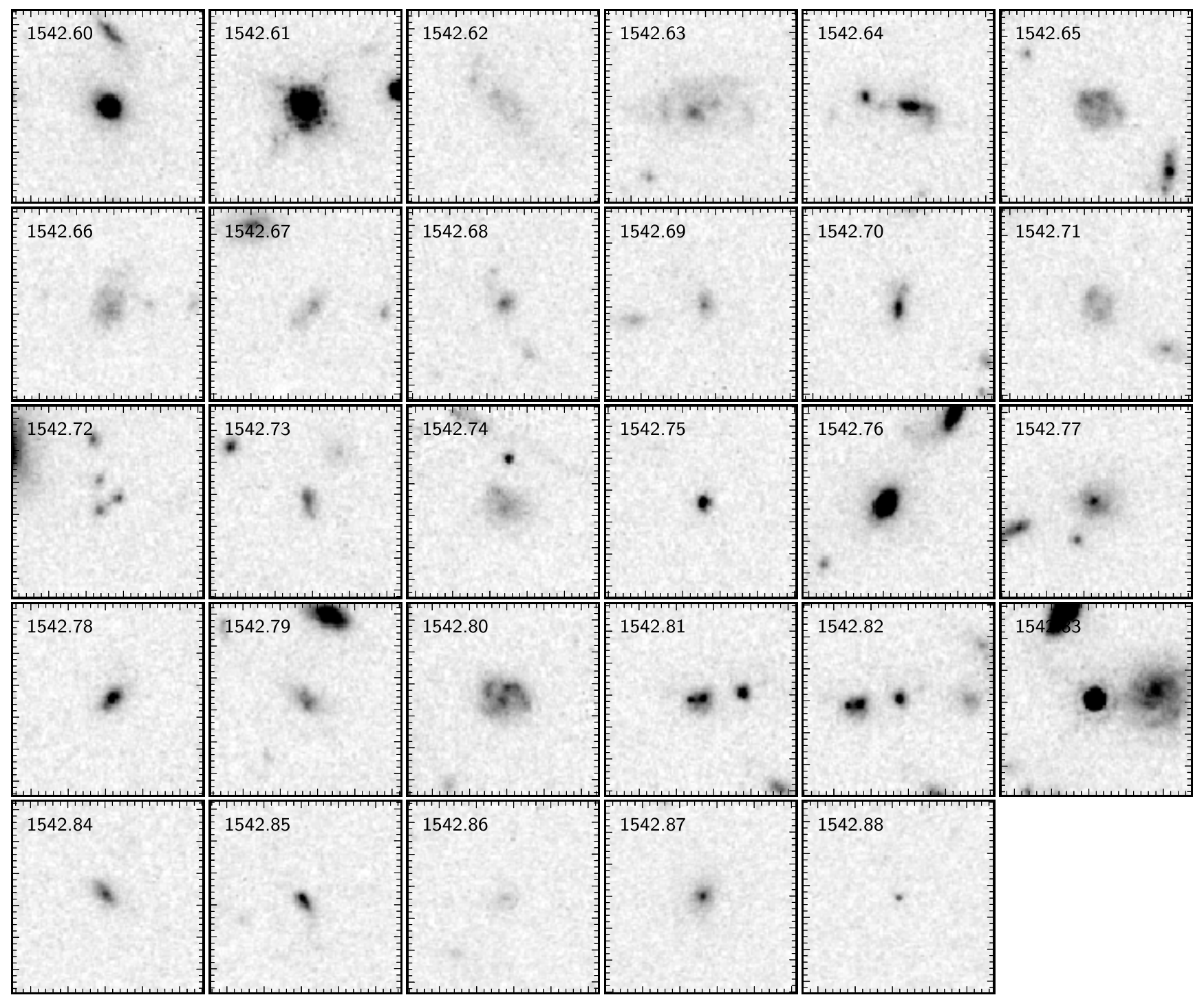}
    \contcaption{}
\end{figure*}

\onecolumn

\begin{landscape}
\begin{longtable}[h]{ccccccccccccc} 
     \caption{All best-fitting parameters in {\tt{SExtractor}} and {\tt{GALFIT}}.}\label{alltable}
     \\ \hline \hline
     ID  &  R. A.  &  Decl. &  Redshift &  $\log(M_\ast$/M$_\odot$)  & $Ks_{\rm best}$  & $H_{\rm best}$ &  $H_{\rm galfit}$ &  $r_{\rm e}$ &  $n$  & Type$^{\rm c}$ \\
        &   (J2000.0)  &  (J2000.0) & $z$ &    & (mag) & (mag)  & (mag) & (\arcsec) &  &  &  &  \\ \hline \endfirsthead
    \multicolumn{13}{c}{continued} \\  \hline \hline
        \endhead
        \hline \endfoot
        \hline \endlastfoot 
1244.01 & 190.888200 & 35.880830 & 2.245 & 10.87$\pm$0.03 & 22.18$\pm$0.05 & 23.48$\pm$0.21 & 23.47$\pm$0.05 & 0.35$\pm$0.03 & 2.56$\pm$0.20 & 3 \\
1244.02 & 190.873101 & 35.885805 & 2.244 & $\cdots$ & 22.21$\pm$0.04 & 24.38$\pm$0.35 & 24.28$\pm$0.01 & 0.22$\pm$0.01 & 0.55$\pm$0.08 & 1 \\
1244.03 & 190.896449 & 35.887899 & 2.255 & 10.81$\pm$0.02 & 21.74$\pm$0.04 & 23.55$\pm$0.22 & 23.39$\pm$0.03 & 0.36$\pm$0.02 & 1.28$\pm$0.10 & 2 \\
1244.04 & 190.871754 & 35.895765 & 2.242 & 10.48$\pm$0.04 & 22.55$\pm$0.07 & 22.84$\pm$0.17 & 22.82$\pm$0.01 & 0.47$\pm$0.01 & 0.84$\pm$0.03 & 3 \\
1244.05 & 190.865368 & 35.899288 & 2.253 & 10.77$\pm$0.03 & 22.08$\pm$0.06 & 22.74$\pm$0.37 & 22.65$\pm$0.01 & 0.58$\pm$0.01 & 0.42$\pm$0.01 & 3 \\
1244.06 & 190.868428 & 35.916471 & 2.213 & 11.04$\pm$0.02 & 21.54$\pm$0.04 & 24.52$\pm$0.35 & 24.71$\pm$0.13 & 0.43$\pm$0.16 & 7.90$\pm$4.55 & 3 \\
1244.07 & 190.885163 & 35.912778 & 2.250 & 11.11$\pm$0.02 & 21.77$\pm$0.03 & 22.66$\pm$0.15 & 22.40$\pm$0.02 & 0.20$\pm$0.01 & 4.59$\pm$0.20 & 1 \\
1244.08 & 190.927770 & 35.914871 & 2.228 & 10.31$\pm$0.06 & 23.12$\pm$0.09 & 23.47$\pm$0.24 & 23.27$\pm$0.02 & 0.40$\pm$0.01 & 1.38$\pm$0.05 & 1 \\
1244.09 & 190.888737 & 35.919665 & 2.220 & 10.79$\pm$0.04 & 22.52$\pm$0.07 & 23.60$\pm$0.24 & 23.24$\pm$0.04 & 0.12$\pm$0.01 & 7.51$\pm$0.96 & 2 \\
1244.10 & 190.910111 & 35.922661 & 2.225 & $\cdots$ & 21.69$\pm$0.04 & 23.25$\pm$0.18 & 22.96$\pm$0.02 & 0.12$\pm$0.01 & 8.00$\pm$0.59 & 2 \\
1244.11 & 190.915332 & 35.927837 & 2.227 & 11.05$\pm$0.02 & 22.08$\pm$0.04 & 22.91$\pm$0.17 & 22.98$\pm$0.01 & 0.23$\pm$0.00 & 1.09$\pm$0.03 & 2 \\
1244.12 & 190.873776 & 35.930775 & 2.247 & 10.31$\pm$0.07 & 23.09$\pm$0.11 & 23.62$\pm$0.22 & 23.50$\pm$0.04 & 0.69$\pm$0.03 & 1.09$\pm$0.06 & 3 \\
1244.13 & 190.874368 & 35.932642 & 2.246 & 10.87$\pm$0.02 & 22.27$\pm$0.04 & 23.23$\pm$0.20 & 23.05$\pm$0.01 & 0.19$\pm$0.00 & 1.19$\pm$0.05 & 2 \\
1244.14 & 190.911658 & 35.936436 & 2.226 & 11.15$\pm$0.02 & 21.80$\pm$0.05 & 22.98$\pm$0.18 & 22.70$\pm$0.02 & 0.68$\pm$0.02 & 1.96$\pm$0.07 & 3 \\
1244.15 & 190.902874 & 35.939224 & 2.230 & 10.94$\pm$0.03 & 21.71$\pm$0.04 & 23.37$\pm$0.20 & 23.33$\pm$0.02 & 0.47$\pm$0.01 & 1.87$\pm$0.07 & 2 \\
1244.16 & 190.861982 & 35.968540 & 2.235 & 10.63$\pm$0.05 & 22.45$\pm$0.08 & 23.48$\pm$0.23 & 23.19$\pm$0.04 & 0.25$\pm$0.02 & 5.42$\pm$0.50 & 2 \\
1244.17 & 190.866592 & 35.971319 & 2.221 & 11.36$\pm$0.01 & 20.56$\pm$0.01 & 21.59$\pm$0.09 & 21.42$\pm$0.01 & 0.10$\pm$0.00 & 8.00$\pm$0.25 & 0 \\
1244.18 & 190.853947 & 35.977385 & 2.235 & 10.51$\pm$0.04 & 22.79$\pm$0.08 & 23.36$\pm$0.22 & 23.24$\pm$0.01 & 0.19$\pm$0.00 & 0.63$\pm$0.04 & 1 \\
1244.19 & 190.850055 & 35.979994 & 2.234 & 10.47$\pm$0.07 & 23.30$\pm$0.14 & 25.01$\pm$0.41 & 24.14$\pm$0.26 & 0.65$\pm$0.42 & 8.00$\pm$2.98 & 4 \\
1244.20 & 191.021577 & 36.079105 & 2.244 & 10.34$\pm$0.06 & 23.26$\pm$0.10 & 23.41$\pm$0.22 & 23.21$\pm$0.02 & 0.44$\pm$0.02 & 1.74$\pm$0.08 & 1 \\
1542.01 & 235.736890 & 38.824496 & 2.238 & 10.50$\pm$0.05 & 22.62$\pm$0.11 & 23.49$\pm$0.22 & 23.25$\pm$0.14 & 0.28$\pm$0.06 & 1.78$\pm$0.53 & 2 \\
1542.02 & 235.741097 & 38.827769 & 2.269 & $\cdots$ & 18.94$\pm$0.02 & 23.32$\pm$0.19 & 22.12$\pm$0.03 & 1.52$\pm$0.05 & 0.87$\pm$0.06 & 4 \\
1542.03 & 235.711379 & 38.828067 & 2.241 & 10.69$\pm$0.04 & 22.27$\pm$0.08 & 22.90$\pm$0.18 & 23.11$\pm$0.04 & 0.55$\pm$0.03 & 0.57$\pm$0.07 & 3 \\
1542.04 & 235.714784 & 38.841468 & 2.241 & 11.00$\pm$0.03 & 21.63$\pm$0.06 & 23.54$\pm$0.24 & 23.43$\pm$0.01 & 0.10$\pm$0.00 & 2.14$\pm$0.19 & 2 \\
1542.05 & 235.720108 & 38.842011 & 2.237 & 10.87$\pm$0.03 & 22.44$\pm$0.07 & 24.24$\pm$0.30 & 23.00$\pm$0.11 & 1.68$\pm$0.33 & 3.51$\pm$0.43 & 4 \\
1542.06 & 235.705981 & 38.866259 & 2.256 & 10.71$\pm$0.03 & 22.10$\pm$0.06 & 23.01$\pm$0.18 & 22.75$\pm$0.02 & 0.27$\pm$0.01 & 3.03$\pm$0.13 & 2 \\
1542.07 & 235.710419 & 38.870823 & 2.242 & 10.56$\pm$0.03 & 22.78$\pm$0.07 & 23.06$\pm$0.19 & 23.00$\pm$0.01 & 0.29$\pm$0.00 & 0.52$\pm$0.02 & 4 \\
1542.08 & 235.707391 & 38.880668 & 2.241 & 10.54$\pm$0.05 & 22.94$\pm$0.11 & 25.04$\pm$0.43 & 23.88$\pm$0.02 & 0.46$\pm$0.01 & 0.58$\pm$0.06 & 4 \\
1542.09 & 235.706498 & 38.889022 & 2.243 & $\cdots$ & 20.77$\pm$0.07 & 22.68$\pm$0.15 & 21.60$\pm$0.26 & 1.63$\pm$0.31 & 0.20$\pm$0.50 & 2 \\
1542.10 & 235.701029 & 38.905092 & 2.238 & 10.89$\pm$0.04 & 22.12$\pm$0.08 & 23.89$\pm$0.26 & 23.20$\pm$0.09 & 0.83$\pm$0.09 & 1.30$\pm$0.15 & 4 \\
1542.11 & 235.688964 & 38.914661 & 2.215 & 10.37$\pm$0.08 & 23.06$\pm$0.15 & 23.89$\pm$0.29 & 23.59$\pm$0.04 & 0.61$\pm$0.03 & 1.82$\pm$0.12 & 2 \\
1542.12 & 235.673164 & 38.947461 & 2.240 & 11.10$\pm$0.02 & 21.75$\pm$0.05 & 22.48$\pm$0.14 & 22.40$\pm$0.01 & 0.38$\pm$0.00 & 0.92$\pm$0.02 & 4 \\
1542.13 & 235.673129 & 38.999895 & 2.227 & 10.19$\pm$0.07 & 23.31$\pm$0.14 & 23.38$\pm$0.21 & 23.25$\pm$0.01 & 0.14$\pm$0.00 & 1.56$\pm$0.08 & 2 \\
1542.14 & 235.670992 & 39.005303 & 2.251 & 10.82$\pm$0.02 & 22.09$\pm$0.05 & 23.14$\pm$0.19 & 22.96$\pm$0.03 & 0.03$\pm$0.00 & 8.00$\pm$1.26 & 2 \\
1542.15 & 235.651288 & 39.005119 & 2.240 & 10.45$\pm$0.05 & 22.69$\pm$0.10 & 23.31$\pm$0.22 & 23.39$\pm$0.02 & 0.54$\pm$0.01 & 0.52$\pm$0.03 & 2 \\
1542.16 & 235.677742 & 39.027920 & 2.206 & 10.80$\pm$0.03 & 22.17$\pm$0.06 & 23.50$\pm$0.22 & 23.00$\pm$0.04 & 0.43$\pm$0.03 & 3.84$\pm$0.23 & 2 \\

1244.21 & 190.931542 & 35.838423 & $\cdots$ & 11.23$\pm$0.02 & 21.59$\pm$0.04 & 22.61$\pm$0.15 & 22.53$\pm$0.01 & 0.41$\pm$0.00 & 0.81$\pm$0.02 & 1 \\
1244.22 & 190.905495 & 35.848431 & $\cdots$ & 10.94$\pm$0.03 & 22.06$\pm$0.07 & 22.98$\pm$0.18 & 22.97$\pm$0.01 & 0.47$\pm$0.01 & 0.92$\pm$0.03 & 2 \\
1244.23 & 190.938369 & 35.849224 & $\cdots$ & 11.48$\pm$0.01 & 20.92$\pm$0.02 & 21.95$\pm$0.11 & 21.81$\pm$0.00 & 0.24$\pm$0.00 & 1.51$\pm$0.02 & 1 \\
1244.24 & 190.922143 & 35.851494 & $\cdots$ & 11.06$\pm$0.02 & 21.98$\pm$0.04 & 23.21$\pm$0.21 & 23.35$\pm$0.01 & 0.38$\pm$0.00 & 0.20$\pm$0.02 & 3 \\
1244.25 & 190.902876 & 35.851714 & $\cdots$ & 10.36$\pm$0.06 & 23.29$\pm$0.11 & 23.82$\pm$0.26 & 23.84$\pm$0.01 & 0.25$\pm$0.00 & 0.20$\pm$0.04 & 3 \\
1244.26 & 190.917185 & 35.852642 & $\cdots$ & 11.33$\pm$0.02 & 21.32$\pm$0.04 & 22.40$\pm$0.14 & 22.25$\pm$0.03 & 0.74$\pm$0.03 & 2.84$\pm$0.09 & 2 \\
1244.27 & 190.905476 & 35.852377 & $\cdots$ & 9.98$\pm$0.10 & 23.52$\pm$0.17 & 23.92$\pm$0.28 & 23.95$\pm$0.01 & 0.27$\pm$0.00 & 0.29$\pm$0.04 & 4 \\
1244.28 & 190.918736 & 35.854295 & $\cdots$ & 11.23$\pm$0.02 & 21.59$\pm$0.05 & 23.30$\pm$0.19 & 22.74$\pm$0.16 & 1.05$\pm$0.25 & 4.08$\pm$0.46 & 2 \\
1244.29 & 190.917671 & 35.853538 & $\cdots$ & 10.86$\pm$0.04 & 22.28$\pm$0.08 & 23.47$\pm$0.23 & 23.07$\pm$0.05 & 0.27$\pm$0.03 & 5.49$\pm$0.54 & 2 \\
1244.30 & 190.918222 & 35.858341 & $\cdots$ & 11.05$\pm$0.02 & 21.92$\pm$0.05 & 22.83$\pm$0.17 & 22.69$\pm$0.01 & 0.28$\pm$0.00 & 1.61$\pm$0.04 & 1 \\
1244.31 & 190.920518 & 35.858688 & $\cdots$ & 11.18$\pm$0.02 & 21.59$\pm$0.04 & 22.90$\pm$0.18 & 22.93$\pm$0.01 & 0.19$\pm$0.00 & 0.92$\pm$0.03 & 3 \\
1244.32 & 190.921027 & 35.858966 & $\cdots$ & 11.59$\pm$0.04 & 20.03$\pm$0.03 & 23.50$\pm$0.23 & 23.43$\pm$0.01 & 0.15$\pm$0.00 & 0.68$\pm$0.04 & 2 \\
1244.33 & 190.914719 & 35.867108 & $\cdots$ & 10.89$\pm$0.03 & 22.38$\pm$0.05 & 23.06$\pm$0.20 & 23.00$\pm$0.01 & 0.69$\pm$0.01 & 0.52$\pm$0.02 & 3 \\
1244.34 & 190.909720 & 35.871428 & $\cdots$ & 10.30$\pm$0.05 & 22.80$\pm$0.09 & 23.31$\pm$0.21 & 23.49$\pm$0.01 & 0.25$\pm$0.00 & 0.41$\pm$0.04 & 3 \\
1244.35 & 190.864540 & 35.871280 & $\cdots$ & 10.47$\pm$0.05 & 23.05$\pm$0.08 & 23.36$\pm$0.42 & 23.26$\pm$0.02 & 0.42$\pm$0.01 & 1.18$\pm$0.05 & 1 \\
1244.36 & 190.898424 & 35.880419 & $\cdots$ & $\cdots$ & 23.00$\pm$0.08 & 24.33$\pm$0.34 & 24.20$\pm$0.01 & 0.14$\pm$0.00 & 0.41$\pm$0.10 & 1 \\
1244.37 & 190.901705 & 35.880067 & $\cdots$ & 10.22$\pm$0.09 & 23.18$\pm$0.13 & 24.00$\pm$0.30 & 24.00$\pm$0.01 & 0.47$\pm$0.01 & 0.24$\pm$0.03 & 2 \\
1244.38 & 190.877336 & 35.893474 & $\cdots$ & $\cdots$ & 22.07$\pm$0.06 & 23.57$\pm$0.25 & 23.42$\pm$0.03 & 0.42$\pm$0.02 & 1.31$\pm$0.09 & 2 \\
1244.39 & 190.877815 & 35.898279 & $\cdots$ & 10.22$\pm$0.06 & 23.06$\pm$0.09 & 23.79$\pm$0.27 & 23.62$\pm$0.08 & 0.13$\pm$0.00 & 1.15$\pm$0.26 & 2 \\
1244.40 & 190.869802 & 35.900108 & $\cdots$ & 10.35$\pm$0.04 & 22.81$\pm$0.06 & 23.38$\pm$0.21 & 23.78$\pm$0.04 & 0.05$\pm$0.00 & 3.80$\pm$0.59 & 2 \\
1244.41 & 190.882531 & 35.905266 & $\cdots$ & 10.32$\pm$0.08 & 23.32$\pm$0.14 & 25.34$\pm$0.49 & 24.91$\pm$0.06 & 0.44$\pm$0.01 & 0.20$\pm$0.03 & 3 \\
1244.42 & 190.927508 & 35.908642 & $\cdots$ & 10.44$\pm$0.08 & 23.26$\pm$0.17 & 24.21$\pm$0.32 & 24.20$\pm$0.03 & 0.17$\pm$0.01 & 3.01$\pm$0.47 & 3 \\
1244.43 & 190.920499 & 35.910929 & $\cdots$ & 10.52$\pm$0.04 & 22.69$\pm$0.06 & 23.50$\pm$0.24 & 23.84$\pm$0.02 & 0.32$\pm$0.01 & 0.69$\pm$0.05 & 3 \\
1244.44 & 190.903619 & 35.918463 & $\cdots$ & 9.82$\pm$0.16 & 24.28$\pm$0.26 & 25.64$\pm$0.66 & 25.53$\pm$0.05 & 0.13$\pm$0.02 & 1.86$\pm$0.96 & 2 \\
1244.45 & 190.913498 & 35.922077 & $\cdots$ & 11.15$\pm$0.02 & 21.75$\pm$0.04 & 23.17$\pm$0.17 & 23.06$\pm$0.01 & 0.26$\pm$0.00 & 1.21$\pm$0.03 & 2 \\
1244.46 & 190.872558 & 35.922177 & $\cdots$ & 10.20$\pm$0.05 & 22.88$\pm$0.07 & 22.98$\pm$0.18 & 22.92$\pm$0.01 & 0.41$\pm$0.01 & 0.76$\pm$0.03 & 1 \\
1244.47 & 190.927969 & 35.926193 & $\cdots$ & 10.25$\pm$0.07 & 23.37$\pm$0.13 & 23.72$\pm$0.25 & 23.53$\pm$0.01 & 0.13$\pm$0.00 & 1.78$\pm$0.14 & 1 \\
1244.48 & 190.856641 & 35.926284 & $\cdots$ & 10.60$\pm$0.05 & 23.06$\pm$0.11 & 23.69$\pm$0.25 & 23.55$\pm$0.01 & 0.59$\pm$0.01 & 0.49$\pm$0.04 & 2 \\
1244.49 & 190.914008 & 35.926802 & $\cdots$ & 11.11$\pm$0.03 & 21.93$\pm$0.06 & 23.46$\pm$0.23 & 23.42$\pm$0.02 & 0.29$\pm$0.01 & 1.13$\pm$0.05 & 2 \\
1244.50 & 190.872876 & 35.933068 & $\cdots$ & 10.55$\pm$0.03 & 22.50$\pm$0.06 & 22.94$\pm$0.18 & 22.78$\pm$0.02 & 0.26$\pm$0.01 & 3.08$\pm$0.14 & 1 \\
1244.51 & 190.874248 & 35.933835 & $\cdots$ & 10.99$\pm$0.03 & 21.74$\pm$0.06 & 23.04$\pm$0.17 & 23.57$\pm$0.02 & 0.34$\pm$0.01 & 0.35$\pm$0.03 & 3 \\
1244.52 & 190.904278 & 35.941551 & $\cdots$ & 10.95$\pm$0.01 & 20.39$\pm$0.01 & 21.37$\pm$0.09 & 21.18$\pm$0.35 & 0.02$\pm$0.01 & 3.74$\pm$2.69 & 0 \\
1244.53 & 190.846262 & 35.947437 & $\cdots$ & $\cdots$ & 21.22$\pm$0.04 & 24.35$\pm$0.00 & 23.02$\pm$0.49 & 1.19$\pm$1.19 & 8.00$\pm$2.91 & 2 \\
1244.54 & 190.845666 & 35.948171 & $\cdots$ & 10.56$\pm$0.04 & 23.21$\pm$0.08 & 24.03$\pm$0.31 & 23.84$\pm$0.02 & 0.34$\pm$0.01 & 1.07$\pm$0.06 & 2 \\
1244.55 & 190.865268 & 35.952369 & $\cdots$ & 9.89$\pm$0.12 & 23.56$\pm$0.19 & 23.87$\pm$0.27 & 23.50$\pm$0.05 & 0.27$\pm$0.03 & 4.56$\pm$0.48 & 2 \\
1244.56 & 190.848995 & 35.952897 & $\cdots$ & 10.69$\pm$0.04 & 22.76$\pm$0.09 & 23.52$\pm$0.24 & 23.22$\pm$0.04 & 0.23$\pm$0.02 & 5.06$\pm$0.46 & 1 \\
1244.57 & 190.838667 & 35.953187 & $\cdots$ & 10.48$\pm$0.04 & 22.46$\pm$0.06 & 22.85$\pm$0.17 & 22.77$\pm$0.01 & 0.34$\pm$0.00 & 1.25$\pm$0.03 & 1 \\
1244.58 & 190.873932 & 35.957972 & $\cdots$ & 10.91$\pm$0.03 & 22.14$\pm$0.06 & 23.33$\pm$0.22 & 23.13$\pm$0.02 & 0.49$\pm$0.01 & 1.30$\pm$0.04 & 1 \\
1244.59 & 190.857448 & 35.964242 & $\cdots$ & 10.76$\pm$0.03 & 21.79$\pm$0.05 & 22.29$\pm$0.13 & 22.16$\pm$0.01 & 0.34$\pm$0.00 & 2.24$\pm$0.05 & 1 \\
1244.60 & 190.862949 & 35.967703 & $\cdots$ & 10.33$\pm$0.06 & 23.10$\pm$0.10 & 23.86$\pm$0.27 & 23.89$\pm$0.01 & 0.20$\pm$0.00 & 0.67$\pm$0.07 & 2 \\
1244.61 & 190.851827 & 35.979160 & $\cdots$ & 10.31$\pm$0.05 & 23.10$\pm$0.09 & 23.55$\pm$0.24 & 23.36$\pm$0.02 & 0.26$\pm$0.01 & 2.81$\pm$0.20 & 2 \\
1244.62 & 190.835762 & 36.014345 & $\cdots$ & 10.85$\pm$0.03 & 22.04$\pm$0.07 & 22.79$\pm$0.17 & 22.96$\pm$0.02 & 0.41$\pm$0.01 & 0.76$\pm$0.03 & 3 \\
1244.63 & 190.831967 & 36.014451 & $\cdots$ & $\cdots$ & 22.60$\pm$0.09 & 23.87$\pm$0.30 & 23.75$\pm$0.02 & 0.45$\pm$0.02 & 0.98$\pm$0.07 & 1 \\
1244.64 & 190.834694 & 36.015956 & $\cdots$ & 10.87$\pm$0.03 & 22.35$\pm$0.06 & 23.25$\pm$0.21 & 23.18$\pm$0.01 & 0.29$\pm$0.00 & 0.62$\pm$0.03 & 1 \\
1244.65 & 190.862015 & 36.022370 & $\cdots$ & 10.62$\pm$0.04 & 22.50$\pm$0.09 & 23.71$\pm$0.24 & 23.39$\pm$0.00 & 0.37$\pm$0.00 & 0.20$\pm$0.00 & 3 \\
1244.66 & 190.865059 & 36.022622 & $\cdots$ & $\cdots$ & 22.62$\pm$0.10 & 23.82$\pm$0.27 & 23.65$\pm$0.02 & 0.15$\pm$0.00 & 2.57$\pm$0.24 & 2 \\
1244.67 & 190.844977 & 36.025253 & $\cdots$ & 10.59$\pm$0.03 & 22.57$\pm$0.05 & 22.99$\pm$0.19 & 22.96$\pm$0.01 & 0.40$\pm$0.00 & 0.31$\pm$0.01 & 2 \\
1244.68 & 190.836299 & 36.029602 & $\cdots$ & 10.83$\pm$0.03 & 22.28$\pm$0.05 & 23.46$\pm$0.21 & 23.41$\pm$0.03 & 0.49$\pm$0.02 & 1.55$\pm$0.07 & 2 \\
1244.69 & 190.849289 & 36.041261 & $\cdots$ & 10.82$\pm$0.02 & 22.19$\pm$0.04 & 23.57$\pm$0.22 & 23.23$\pm$0.01 & 0.63$\pm$0.01 & 0.53$\pm$0.03 & 2 \\
1244.70 & 190.841886 & 36.041627 & $\cdots$ & 10.08$\pm$0.11 & 23.49$\pm$0.15 & 23.97$\pm$0.27 & 24.10$\pm$0.02 & 0.29$\pm$0.01 & 1.22$\pm$0.07 & 2 \\
1244.71 & 190.833760 & 36.044398 & $\cdots$ & 10.54$\pm$0.06 & 23.05$\pm$0.12 & 25.17$\pm$0.53 & 24.05$\pm$0.40 & 1.56$\pm$1.47 & 8.00$\pm$3.44 & 1 \\
1244.72 & 190.845685 & 36.046986 & $\cdots$ & 11.04$\pm$0.02 & 21.86$\pm$0.04 & 22.56$\pm$0.15 & 22.36$\pm$0.02 & 0.10$\pm$0.00 & 8.00$\pm$0.57 & 2 \\
1244.73 & 190.834339 & 36.053715 & $\cdots$ & 10.56$\pm$0.05 & 22.55$\pm$0.07 & 23.36$\pm$0.22 & 23.33$\pm$0.02 & 0.26$\pm$0.01 & 1.71$\pm$0.09 & 1 \\
1244.74 & 190.822668 & 36.062741 & $\cdots$ & $\cdots$ & 19.27$\pm$0.02 & 23.52$\pm$0.22 & 23.50$\pm$0.27 & 0.69$\pm$0.36 & 2.89$\pm$1.61 & 2 \\
1244.75 & 191.018688 & 36.061936 & $\cdots$ & 10.28$\pm$0.08 & 23.27$\pm$0.12 & 23.83$\pm$0.26 & 23.72$\pm$0.01 & 0.22$\pm$0.00 & 0.62$\pm$0.05 & 1 \\
1244.76 & 191.005548 & 36.064456 & $\cdots$ & 9.97$\pm$0.16 & 23.18$\pm$0.17 & 24.39$\pm$0.35 & 24.22$\pm$0.09 & 0.35$\pm$0.06 & 3.87$\pm$0.74 & 2 \\
1244.77 & 190.997276 & 36.067849 & $\cdots$ & 10.38$\pm$0.06 & 22.75$\pm$0.12 & 22.99$\pm$0.18 & 22.95$\pm$0.01 & 0.35$\pm$0.00 & 0.64$\pm$0.03 & 1 \\
1244.78 & 190.855419 & 36.068327 & $\cdots$ & 10.16$\pm$0.09 & 23.27$\pm$0.17 & 23.72$\pm$0.26 & 23.63$\pm$0.02 & 0.19$\pm$0.01 & 1.95$\pm$0.17 & 2 \\
1244.79 & 190.981504 & 36.069242 & $\cdots$ & $\cdots$ & 24.20$\pm$0.26 & 25.11$\pm$0.52 & 25.01$\pm$0.02 & 0.19$\pm$0.01 & 0.20$\pm$0.11 & 1 \\
1244.80 & 190.843856 & 36.070453 & $\cdots$ & 9.87$\pm$0.08 & 23.28$\pm$0.14 & 23.21$\pm$0.20 & 23.10$\pm$0.01 & 0.25$\pm$0.00 & 0.98$\pm$0.04 & 1 \\
1244.81 & 190.988484 & 36.072096 & $\cdots$ & 10.59$\pm$0.04 & 22.54$\pm$0.08 & 22.91$\pm$0.18 & 22.80$\pm$0.01 & 0.39$\pm$0.01 & 0.89$\pm$0.02 & 1 \\
1244.82 & 191.013139 & 36.076198 & $\cdots$ & 10.21$\pm$0.06 & 22.95$\pm$0.09 & 24.23$\pm$0.29 & 23.56$\pm$0.03 & 0.51$\pm$0.03 & 2.11$\pm$0.16 & 2 \\
1244.83 & 190.985994 & 36.076362 & $\cdots$ & 10.46$\pm$0.05 & 22.50$\pm$0.08 & 23.25$\pm$0.21 & 23.07$\pm$0.02 & 0.53$\pm$0.02 & 2.09$\pm$0.08 & 2 \\
1244.84 & 191.020413 & 36.081577 & $\cdots$ & 10.26$\pm$0.06 & 23.08$\pm$0.11 & 23.09$\pm$0.19 & 22.85$\pm$0.01 & 0.31$\pm$0.01 & 2.72$\pm$0.10 & 1 \\
1244.85 & 191.004787 & 36.082345 & $\cdots$ & 9.81$\pm$0.12 & 23.87$\pm$0.15 & 23.67$\pm$0.25 & 23.55$\pm$0.01 & 0.34$\pm$0.01 & 0.95$\pm$0.05 & 2 \\
1542.17 & 235.816701 & 38.792986 & $\cdots$ & 10.82$\pm$0.02 & 21.40$\pm$0.03 & 21.52$\pm$0.09 & 21.46$\pm$0.01 & 0.36$\pm$0.00 & 1.23$\pm$0.02 & 1 \\
1542.18 & 235.831037 & 38.797937 & $\cdots$ & 10.94$\pm$0.03 & 21.93$\pm$0.06 & 22.79$\pm$0.17 & 22.58$\pm$0.02 & 0.86$\pm$0.03 & 2.52$\pm$0.10 & 1 \\
1542.19 & 235.822202 & 38.798663 & $\cdots$ & 10.26$\pm$0.05 & 23.36$\pm$0.10 & 23.49$\pm$0.23 & 23.26$\pm$0.02 & 0.24$\pm$0.01 & 2.38$\pm$0.15 & 2 \\
1542.20 & 235.829372 & 38.799283 & $\cdots$ & 10.55$\pm$0.06 & 23.08$\pm$0.14 & 23.57$\pm$0.25 & 23.33$\pm$0.02 & 0.44$\pm$0.02 & 1.46$\pm$0.07 & 1 \\
1542.21 & 235.831365 & 38.799592 & $\cdots$ & 10.78$\pm$0.04 & 22.32$\pm$0.08 & 22.96$\pm$0.19 & 22.87$\pm$0.01 & 0.56$\pm$0.01 & 0.76$\pm$0.03 & 3 \\
1542.22 & 235.818707 & 38.805318 & $\cdots$ & 11.18$\pm$0.02 & 21.73$\pm$0.05 & 22.85$\pm$0.17 & 22.73$\pm$0.02 & 0.50$\pm$0.02 & 2.11$\pm$0.07 & 1 \\
1542.23 & 235.827769 & 38.807361 & $\cdots$ & $\cdots$ & 22.57$\pm$0.07 & 23.64$\pm$0.25 & 23.57$\pm$0.02 & 0.25$\pm$0.01 & 1.61$\pm$0.12 & 1 \\
1542.24 & 235.730143 & 38.814212 & $\cdots$ & 10.87$\pm$0.03 & 22.18$\pm$0.06 & 22.91$\pm$0.18 & 22.93$\pm$0.01 & 0.32$\pm$0.01 & 1.05$\pm$0.04 & 2 \\
1542.25 & 235.842034 & 38.814018 & $\cdots$ & 10.43$\pm$0.06 & 23.05$\pm$0.11 & 23.62$\pm$0.24 & 23.53$\pm$0.01 & 0.34$\pm$0.01 & 0.95$\pm$0.05 & 2 \\
1542.26 & 235.852340 & 38.814727 & $\cdots$ & 10.63$\pm$0.03 & 22.34$\pm$0.06 & 22.63$\pm$0.15 & 22.51$\pm$0.01 & 0.64$\pm$0.01 & 0.69$\pm$0.02 & 3 \\
1542.27 & 235.721645 & 38.818959 & $\cdots$ & $\cdots$ & 23.19$\pm$0.10 & 23.74$\pm$0.27 & 23.62$\pm$0.01 & 0.36$\pm$0.01 & 0.59$\pm$0.04 & 2 \\
1542.28 & 235.833842 & 38.819854 & $\cdots$ & 10.10$\pm$0.06 & 23.08$\pm$0.10 & 23.20$\pm$0.20 & 23.11$\pm$0.01 & 0.26$\pm$0.00 & 0.76$\pm$0.03 & 2 \\
1542.29 & 235.837325 & 38.824165 & $\cdots$ & 10.06$\pm$0.06 & 22.85$\pm$0.10 & 23.95$\pm$0.27 & 23.75$\pm$0.02 & 0.12$\pm$0.01 & 2.34$\pm$0.38 & 2 \\
1542.30 & 235.834175 & 38.824990 & $\cdots$ & 11.28$\pm$0.01 & 21.28$\pm$0.03 & 22.25$\pm$0.13 & 22.15$\pm$0.14 & 0.36$\pm$0.09 & 3.77$\pm$1.26 & 2 \\
1542.31 & 235.838175 & 38.824298 & $\cdots$ & 10.14$\pm$0.06 & 23.19$\pm$0.10 & 23.68$\pm$0.25 & 23.63$\pm$0.01 & 0.21$\pm$0.00 & 0.42$\pm$0.04 & 4 \\
1542.32 & 235.840612 & 38.825680 & $\cdots$ & 9.87$\pm$0.10 & 23.39$\pm$0.15 & 23.44$\pm$0.23 & 22.88$\pm$0.07 & 0.57$\pm$0.10 & 6.87$\pm$0.77 & 3 \\
1542.33 & 235.715899 & 38.826737 & $\cdots$ & 10.90$\pm$0.02 & 22.37$\pm$0.05 & 22.94$\pm$0.18 & 22.92$\pm$0.02 & 0.35$\pm$0.01 & 0.29$\pm$0.05 & 3 \\
1542.34 & 235.712866 & 38.830824 & $\cdots$ & 10.54$\pm$0.04 & 22.61$\pm$0.08 & 23.69$\pm$0.24 & 23.53$\pm$0.01 & 0.14$\pm$0.00 & 1.40$\pm$0.09 & 2 \\
1542.35 & 235.713168 & 38.832283 & $\cdots$ & 11.45$\pm$0.06 & 19.98$\pm$0.02 & 23.84$\pm$0.28 & 23.79$\pm$0.01 & 0.23$\pm$0.00 & 0.20$\pm$0.04 & 4 \\
1542.36 & 235.713166 & 38.832737 & $\cdots$ & 10.93$\pm$0.03 & 21.81$\pm$0.06 & 23.30$\pm$0.21 & 23.06$\pm$0.04 & 0.32$\pm$0.02 & 3.65$\pm$0.27 & 2 \\
1542.37 & 235.742718 & 38.833981 & $\cdots$ & 10.12$\pm$0.09 & 23.38$\pm$0.20 & 23.19$\pm$0.21 & 23.10$\pm$0.01 & 0.58$\pm$0.01 & 0.51$\pm$0.03 & 3 \\
1542.38 & 235.846127 & 38.835442 & $\cdots$ & 10.76$\pm$0.03 & 22.21$\pm$0.08 & 23.06$\pm$0.17 & 22.94$\pm$0.01 & 0.38$\pm$0.01 & 0.90$\pm$0.03 & 4 \\
1542.39 & 235.712409 & 38.836037 & $\cdots$ & $\cdots$ & 23.22$\pm$0.12 & 24.26$\pm$0.34 & 24.13$\pm$0.02 & 0.37$\pm$0.01 & 0.93$\pm$0.09 & 3 \\
1542.40 & 235.734801 & 38.837216 & $\cdots$ & 9.92$\pm$0.07 & 23.05$\pm$0.10 & 23.06$\pm$0.18 & 23.12$\pm$0.01 & 0.05$\pm$0.00 & 3.05$\pm$0.34 & 2 \\
1542.41 & 235.720914 & 38.839631 & $\cdots$ & 10.58$\pm$0.06 & 22.98$\pm$0.13 & 24.33$\pm$0.35 & 24.25$\pm$0.05 & 0.16$\pm$0.01 & 3.95$\pm$0.79 & 2 \\
1542.42 & 235.843548 & 38.844702 & $\cdots$ & 10.96$\pm$0.03 & 22.16$\pm$0.06 & 23.26$\pm$0.22 & 23.12$\pm$0.02 & 0.65$\pm$0.02 & 1.68$\pm$0.06 & 4 \\
1542.43 & 235.852174 & 38.844649 & $\cdots$ & 10.25$\pm$0.06 & 23.29$\pm$0.11 & 23.43$\pm$0.22 & 23.29$\pm$0.01 & 0.25$\pm$0.00 & 1.16$\pm$0.05 & 1 \\
1542.44 & 235.853698 & 38.845456 & $\cdots$ & 10.73$\pm$0.03 & 22.40$\pm$0.06 & 22.84$\pm$0.17 & 22.13$\pm$0.07 & 1.66$\pm$0.26 & 7.81$\pm$0.56 & 1 \\
1542.45 & 235.720987 & 38.845898 & $\cdots$ & 10.22$\pm$0.06 & 23.29$\pm$0.12 & 23.42$\pm$0.22 & 23.32$\pm$0.01 & 0.28$\pm$0.01 & 1.43$\pm$0.07 & 1 \\
1542.46 & 235.853063 & 38.846656 & $\cdots$ & 10.57$\pm$0.04 & 22.93$\pm$0.10 & 23.68$\pm$0.25 & 23.52$\pm$0.02 & 0.21$\pm$0.01 & 2.61$\pm$0.21 & 2 \\
1542.47 & 235.838158 & 38.847944 & $\cdots$ & 10.43$\pm$0.05 & 23.12$\pm$0.11 & 24.19$\pm$0.29 & 24.10$\pm$0.02 & 0.18$\pm$0.01 & 0.97$\pm$0.08 & 2 \\
1542.48 & 235.707328 & 38.852077 & $\cdots$ & 10.24$\pm$0.07 & 23.32$\pm$0.16 & 23.27$\pm$0.31 & 22.98$\pm$0.03 & 0.65$\pm$0.03 & 1.89$\pm$0.10 & 1 \\
1542.49 & 235.713261 & 38.853924 & $\cdots$ & 10.30$\pm$0.05 & 23.32$\pm$0.10 & 23.64$\pm$0.24 & 23.47$\pm$0.02 & 0.18$\pm$0.00 & 2.00$\pm$0.13 & 1 \\
1542.50 & 235.717072 & 38.857236 & $\cdots$ & 10.50$\pm$0.06 & 23.03$\pm$0.13 & 23.18$\pm$0.20 & 23.03$\pm$0.02 & 0.67$\pm$0.02 & 0.70$\pm$0.04 & 1 \\
1542.51 & 235.835606 & 38.858636 & $\cdots$ & 10.83$\pm$0.03 & 22.36$\pm$0.07 & 22.97$\pm$0.20 & 23.28$\pm$0.03 & 0.68$\pm$0.02 & 0.82$\pm$0.04 & 4 \\
1542.52 & 235.718255 & 38.859269 & $\cdots$ & 11.00$\pm$0.02 & 22.22$\pm$0.05 & 23.17$\pm$0.20 & 23.11$\pm$0.01 & 0.46$\pm$0.00 & 0.25$\pm$0.02 & 3 \\
1542.53 & 235.832316 & 38.861574 & $\cdots$ & 10.78$\pm$0.04 & 22.52$\pm$0.08 & 23.51$\pm$0.24 & 23.33$\pm$0.06 & 0.75$\pm$0.06 & 2.43$\pm$0.19 & 3 \\
1542.54 & 235.831824 & 38.861987 & $\cdots$ & 10.99$\pm$0.03 & 22.12$\pm$0.07 & 22.98$\pm$0.19 & 22.77$\pm$0.01 & 0.73$\pm$0.01 & 1.35$\pm$0.04 & 2 \\
1542.55 & 235.819980 & 38.868539 & $\cdots$ & 10.99$\pm$0.03 & 22.28$\pm$0.06 & 23.26$\pm$0.20 & 23.05$\pm$0.02 & 0.21$\pm$0.01 & 4.05$\pm$0.27 & 1 \\
1542.56 & 235.824425 & 38.865864 & $\cdots$ & $\cdots$ & 19.29$\pm$0.02 & 23.45$\pm$0.23 & 23.48$\pm$0.04 & 0.20$\pm$0.01 & 1.26$\pm$0.25 & 2 \\
1542.57 & 235.822256 & 38.865524 & $\cdots$ & $\cdots$ & 21.49$\pm$0.09 & 23.02$\pm$0.17 & 22.99$\pm$0.01 & 0.38$\pm$0.00 & 0.50$\pm$0.03 & 2 \\
1542.58 & 235.738981 & 38.869918 & $\cdots$ & 10.46$\pm$0.04 & 22.91$\pm$0.09 & 23.21$\pm$0.20 & 23.17$\pm$0.01 & 0.52$\pm$0.01 & 0.40$\pm$0.02 & 2 \\
1542.59 & 235.695822 & 38.893258 & $\cdots$ & 11.01$\pm$0.02 & 21.91$\pm$0.05 & 23.29$\pm$0.19 & 23.52$\pm$0.05 & 0.54$\pm$0.01 & 0.20$\pm$0.03 & 3 \\
1542.60 & 235.754135 & 38.894294 & $\cdots$ & 11.56$\pm$0.01 & 20.81$\pm$0.02 & 21.51$\pm$0.09 & 21.26$\pm$0.00 & 0.22$\pm$0.00 & 4.25$\pm$0.05 & 1 \\
1542.61 & 235.695918 & 38.895108 & $\cdots$ & 11.12$\pm$0.01 & 19.42$\pm$0.01 & 19.83$\pm$0.04 & 19.75$\pm$0.04 & 0.02$\pm$0.00 & 7.63$\pm$0.82 & 0 \\
1542.62 & 235.692814 & 38.895989 & $\cdots$ & $\cdots$ & 22.02$\pm$0.08 & 24.91$\pm$0.41 & 25.08$\pm$0.72 & 0.55$\pm$0.96 & 8.00$\pm$7.80 & 3 \\
1542.63 & 235.863511 & 38.895363 & $\cdots$ & 10.41$\pm$0.07 & 23.61$\pm$0.17 & 22.92$\pm$0.16 & 23.28$\pm$0.05 & 0.42$\pm$0.02 & 1.63$\pm$0.09 & 1 \\
1542.64 & 235.738889 & 38.903147 & $\cdots$ & 11.00$\pm$0.02 & 21.65$\pm$0.05 & 22.38$\pm$0.14 & 22.28$\pm$0.01 & 0.53$\pm$0.01 & 1.72$\pm$0.03 & 2 \\
1542.65 & 235.769200 & 38.906334 & $\cdots$ & 10.93$\pm$0.03 & 21.86$\pm$0.06 & 22.91$\pm$0.17 & 22.75$\pm$0.01 & 0.50$\pm$0.00 & 0.20$\pm$0.01 & 4 \\
1542.66 & 235.757562 & 38.908424 & $\cdots$ & 10.72$\pm$0.03 & 22.43$\pm$0.06 & 22.91$\pm$0.19 & 22.87$\pm$0.01 & 0.58$\pm$0.01 & 0.59$\pm$0.02 & 2 \\
1542.67 & 235.760229 & 38.909677 & $\cdots$ & 10.78$\pm$0.04 & 22.57$\pm$0.10 & 23.80$\pm$0.28 & 23.71$\pm$0.01 & 0.54$\pm$0.01 & 0.27$\pm$0.03 & 2 \\
1542.68 & 235.747734 & 38.910716 & $\cdots$ & 10.61$\pm$0.04 & 22.82$\pm$0.10 & 23.57$\pm$0.24 & 23.30$\pm$0.02 & 0.36$\pm$0.01 & 2.18$\pm$0.10 & 1 \\
1542.69 & 235.766410 & 38.916148 & $\cdots$ & $\cdots$ & 22.92$\pm$0.12 & 23.81$\pm$0.28 & 23.49$\pm$0.04 & 0.47$\pm$0.03 & 2.45$\pm$0.18 & 2 \\
1542.70 & 235.689052 & 38.918011 & $\cdots$ & 10.42$\pm$0.06 & 22.78$\pm$0.13 & 23.06$\pm$0.19 & 22.91$\pm$0.04 & 0.42$\pm$0.03 & 1.65$\pm$0.17 & 1 \\
1542.71 & 235.741964 & 38.919411 & $\cdots$ & 10.83$\pm$0.03 & 22.40$\pm$0.07 & 23.79$\pm$0.24 & 23.33$\pm$0.02 & 0.44$\pm$0.01 & 0.36$\pm$0.02 & 4 \\
1542.72 & 235.760582 & 38.922065 & $\cdots$ & $\cdots$ & 22.38$\pm$0.09 & 24.89$\pm$0.35 & 24.90$\pm$0.03 & 0.14$\pm$0.01 & 1.23$\pm$0.29 & 2 \\
1542.73 & 235.680085 & 38.930766 & $\cdots$ & 10.63$\pm$0.04 & 22.82$\pm$0.08 & 23.63$\pm$0.25 & 23.55$\pm$0.01 & 0.32$\pm$0.00 & 0.41$\pm$0.03 & 2 \\
1542.74 & 235.743599 & 38.931790 & $\cdots$ & 11.00$\pm$0.02 & 21.90$\pm$0.05 & 22.81$\pm$0.31 & 22.75$\pm$0.01 & 0.55$\pm$0.01 & 0.50$\pm$0.02 & 1 \\
1542.75 & 235.684373 & 38.932120 & $\cdots$ & 9.32$\pm$0.18 & 24.12$\pm$0.28 & 23.42$\pm$0.22 & 23.34$\pm$0.00 & 0.14$\pm$0.00 & 0.20$\pm$0.05 & 1 \\
1542.76 & 235.706072 & 38.940894 & $\cdots$ & 11.51$\pm$0.01 & 20.49$\pm$0.01 & 21.25$\pm$0.08 & 21.08$\pm$0.00 & 0.19$\pm$0.00 & 3.95$\pm$0.05 & 2 \\
1542.77 & 235.695391 & 38.937377 & $\cdots$ & 10.38$\pm$0.04 & 22.79$\pm$0.07 & 22.92$\pm$0.17 & 22.74$\pm$0.01 & 0.28$\pm$0.01 & 2.31$\pm$0.08 & 1 \\
1542.78 & 235.670964 & 38.953147 & $\cdots$ & 10.41$\pm$0.07 & 23.05$\pm$0.15 & 23.23$\pm$0.21 & 23.19$\pm$0.02 & 0.41$\pm$0.01 & 0.82$\pm$0.05 & 3 \\
1542.79 & 235.694625 & 38.958472 & $\cdots$ & 11.12$\pm$0.02 & 21.60$\pm$0.05 & 23.02$\pm$0.17 & 22.86$\pm$0.04 & 0.51$\pm$0.01 & 0.99$\pm$0.03 & 3 \\
1542.80 & 235.700570 & 38.959961 & $\cdots$ & 11.90$\pm$0.02 & 19.14$\pm$0.02 & 24.72$\pm$0.41 & 22.99$\pm$0.24 & 1.15$\pm$0.65 & 8.00$\pm$2.12 & 4 \\
1542.81 & 235.700119 & 38.960014 & $\cdots$ & 11.62$\pm$0.03 & 19.94$\pm$0.02 & 23.45$\pm$0.23 & 23.26$\pm$0.03 & 0.12$\pm$0.01 & 5.64$\pm$0.70 & 2 \\
1542.82 & 235.703679 & 38.960586 & $\cdots$ & 10.00$\pm$0.02 & 20.48$\pm$0.04 & 20.82$\pm$0.07 & 20.70$\pm$0.05 & 0.02$\pm$0.00 & 3.56$\pm$0.93 & 2 \\
1542.83 & 235.646645 & 39.002300 & $\cdots$ & 10.53$\pm$0.04 & 22.97$\pm$0.07 & 23.39$\pm$0.22 & 23.29$\pm$0.01 & 0.34$\pm$0.01 & 0.92$\pm$0.04 & 1 \\
1542.84 & 235.678297 & 39.007564 & $\cdots$ & 10.00$\pm$0.08 & 23.40$\pm$0.12 & 23.45$\pm$0.22 & 23.36$\pm$0.01 & 0.25$\pm$0.00 & 1.06$\pm$0.05 & 1 \\
1542.85 & 235.640541 & 39.009079 & $\cdots$ & $\cdots$ & 22.75$\pm$0.09 & 24.67$\pm$0.43 & 24.51$\pm$0.03 & 0.34$\pm$0.02 & 0.90$\pm$0.11 & 2 \\
1542.86 & 235.643524 & 39.010000 & $\cdots$ & 10.89$\pm$0.04 & 22.39$\pm$0.09 & 23.12$\pm$0.20 & 22.83$\pm$0.03 & 0.51$\pm$0.02 & 2.67$\pm$0.11 & 1 \\
1542.87 & 235.658678 & 39.025457 & $\cdots$ & $\cdots$ & 24.02$\pm$0.24 & 25.66$\pm$0.58 & 25.47$\pm$0.06 & 0.02$\pm$0.02 & 4.35$\pm$6.06 & 1 \\
\end{longtable}
$^{\rm a}$ The BEST magnitudes in {\textsc{SExtractor}}. $^{\rm b}$ The {\textsc{GALFIT}} model magnitudes.
$^{\rm c}$ Type=0, is a QSO; Type=1, is an isolated galaxies; Type=2, has close neighbors within 30\,kpc; Type=3, possess merger features within 5\,kpc; Type=4, merging galaxy with close companions.
\end{landscape}


\bsp 
\label{lastpage}
\end{document}